\documentclass[twocolumn]{autart}

\usepackage{natbib} 
\usepackage{amssymb,amsmath,amsfonts}
\usepackage{latexsym,graphicx,color}
\usepackage{mathrsfs} 

\usepackage[linguistics]{forest}
\usepackage{standalone,multirow,arydshln}
\usepackage{rotating}
\usepackage{clrscode3e}
\usepackage{xparse,mathtools}
\usepackage{textcomp}
\usepackage{arydshln}

\allowdisplaybreaks[4]

\DeclareMathOperator*{\argminA}{arg\,min}
\def\cal{\mathcal} 

\def\allwords{\mbox{$X^\ast$}}

\definecolor{Light}{gray}{0.85}

\def\abs#1{\left\vert #1 \right\vert}
\def\allpolyx0degn{\mbox{$P_n$}}

\def\allseries{\mbox{$\re\langle\langle X \rangle\rangle$}}

\def\allseriesell{\mbox{$\re^{\ell} \langle\langle X \rangle\rangle$}}

\def\bull{\rule{0.08in}{0.08in}} 

\def\card{{\rm card}}

\newcommand{\comment}[1]{} 

\def\diag{{\rm diag}}

\def\eqref#1{(\ref{#1})} 

\def\expup{{\rm e}}

\def\nat{{\mathbb N}} 
\def\norm#1{\left\Vert#1\right\Vert}

\def\openbull{\framebox[0.08in][c]{$\;$}} 

\def\re{{\mathbb R}} 

\def\vec{{\rm vec}}

\def\begals{\[\begin{aligned}}
\def\endals{\end{aligned}\]}
\def\begce{\begin{center}}
\def\endce{\end{center}}
\def\begar{\begin{array}}
\def\endar{\end{array}}
\def\begeq{\begin{equation}}
\def\endeq{\end{equation}}
\def\begdi{\begin{displaymath}}
\def\enddi{\end{displaymath}}
\def\begdis{\begin{eqnarray*}}
\def\enddis{\end{eqnarray*}}
\def\begeqa{\begin{eqnarray}}
\def\endeqa{\end{eqnarray}}
\def\begdes{\begin{description}}
\def\enddes{\end{description}}
\def\begit{\begin{itemize}}
\def\endit{\end{itemize}}
\def\begen{\begin{enumerate}}
\def\enden{\end{enumerate}}
\def\beglar{\left[\begin{array}}
\def\endrar{\end{array}\right]}
\def\begle{\begin{mylemma}}
\def\endle{\end{mylemma}}
\def\begde{\begin{mydefinition}}
\def\endde{\end{mydefinition}}
\def\begth{\begin{mytheorem}}
\def\endth{\end{mytheorem}}
\def\begco{\begin{mycorollary}}
\def\endco{\end{mycorollary}}
\def\begprop{\begin{myproposition}}
\def\endprop{\end{myproposition}}
\def\begex{\begin{myexample}}
\def\endex{\hfill\openbull \end{myexample} \vspace*{0.15in}}
\def\begexer{\begin{myexercise}}
\def\endexer{\end{myexercise}}

\def\begres{\noindent{\bf Remarks}:\begin{enumerate}}
\def\endres{\end{enumerate} \par}
\def\begpr{\noindent{\em Proof:}$\;\;$}
\def\endpr{\hfill\bull \vspace*{0.15in}}
\def\begtab{\begin{tabular}}
\def\endtab{\end{tabular}}
\def\rref#1{(\ref{#1})}

\def\begle{\begin{lem}}
\def\endle{\end{lem}}
\def\begde{\begin{defn}}
\def\endde{\end{defn}}
\def\begth{\begin{thm}}
\def\endth{\end{thm}}
\def\begco{\begin{cor}}
\def\endco{\end{cor}}
\def\begex{\begin{exmp}\rm}
\def\endex{\end{exmp}}
\def\begpr{\begin{pf}}
\def\endpr{\end{pf}}

\begin{document}

\begin{frontmatter}

\title{Combining Learning and Model Based Control \\
via Discrete-Time Chen-Fliess Series}

\author[First]{W.~Steven Gray\thanksref{corrauthor}}
\author[Second]{G.~S. Venkatesh}
\author[Third]{Luis A.~Duffaut Espinosa}

\thanks[corrauthor]{Corresponding author}

\address[First]{Department of Electrical and Computer Engineering,
Old Dominion University, Norfolk, VA 23529, USA (email: sgray@odu.edu)}
\address[Second]{Department of Electrical and Computer Engineering,
Old Dominion University, Norfolk, VA 23529, USA (email: gsvenky89@gmail.com)}
\address[Third]{Department of Electrical and
Biomedical Engineering, University of Vermont, Burlington, Vermont 05405 USA (email: lduffaut@uvm.edu)}

\begin{abstract} 
A learning control system is presented suitable for control affine nonlinear plants based on
discrete-time Chen-Fliess series and capable of
incorporating knowledge of a given physical model.
The
underlying noncommutative algebraic and combinatorial structures needed to realize the multivariable case are also described.
The method is demonstrated using a two-input, two-output Lotka-Volterra system.
\end{abstract}

\begin{keyword}
Nonlinear systems, learning control, adaptive control, Chen-Fliess series
\end{keyword}

\end{frontmatter}

\section{Introduction}

The central attraction of applying learning/adaptive data science to control is the ability to learn and generalize plant dynamics from partial input-output data
in order to react properly to new situations. This can be done off-line via a training phase and/or online during
closed-loop operation. The most widely used methods at present are based on artificial neural networks (ANNs), reinforcement learning control,
and local adaptive control.

Most of the modern ANN approaches to learning control have their origins in the work of
McCulloch and Pitts in the 1950's and is based on the computational capabilities of networks of individual units called {\em neurons} \citep{McCulloch-Pitts_43}. The approach was further developed by Rosenblatt to produce what is now known as {\em perceptron multilayer feedforward nets} \citep{Rosenblatt_62}. At its core, an ANN in learning control realizes a static map parameterized by a structured set of real parameters that operates on an input signal  \citep{Hunt-et-al_92}.
These parameters are adjusted based on some learning strategy such as backpropagation. The family of mappings in the controller is assumed to be sufficiently rich to represent a wide variety of potential control laws.
But usually only simulation based justifications are possible. A long standing criticism is that there are few theoretical results strongly linking the properties of ANNs to the learning/adaptive control problem \citep{Polycarpou-Ioannou_92}.
Adding dynamics to ANNs to form recurrent neural networks (RNNs) was a natural step in the development of learning methodologies in control. The hope here was that they would better approximate dynamic input-output behaviors \citep{Baldi-Hornik_96,Jin-etal_95,Kambhampati-etal_00,Schaefer-Zimmermann_06,Xiao-Dong_2005}.
They also permit online learning as the network's state evolves over time in response to an applied input much like the state of the plant. But again a major drawback
concerning RNNs is that they lack theoretical support for the intended application of control. One attempt to address this issue is the use of
input convex neural networks for optimal control \citep{Chen-etal_19}.
They take advantage of recent progress in deep learning optimization, but have significant computational overhead. Hence, they are not suitable for every control application.

Reinforcement learning control is based on the classical Lyapunov/Bellman function. Simply stated, the aim is to find the proper control action with respect to an
overall long term objective. A dynamic programming paradigm is used to minimize at each time instant a quantity that measures the overall {\em goodness} of a given state.
This is classically known in optimal control as the Bellman value function or a Lyapunov energy function. The {\em actor-critic approach}
relies on the principle of assigning a cost to
a given state and/or a proposed action in a way that this cost becomes a good predictor of eventual long term outcomes
\citep{Lewis-Vrabie_2009,Lewis-Vrabie-Syrmos_2012,Lewis-et-al_2012,Vrabie-Lewis_2009,Vrabie-et-al_2009}. It originated from specializing
the work of Barto, who first introduced the so called {\em adaptive critic} \citep{Barto_90}.  Lewis et al.\ provided a more systematic
version of the idea using a more suitably posed goal. This work is also related
to that described in \cite{Mendel-Fu_70,Narendra-Thathachar_89}. Specifically, a parametric form of a Lyapunov function (the {\em critic system})
is given and one attempts to fit parameters for the Lyapunov function and the proposed feedback law simultaneously. This is done by adjusting the
parameters after a training event (occurring on a different time scale) via a steepest descent step.

The local adaptive control methodology is based on operating the system about a set of predetermined operating points for which robust
controllers are designed using a corresponding set of linear models \citep{Astrom-Wittenmark_94,Slotine-Li_91}. The learning consists of building an association between the current state and the appropriate controller. This is conceptually a variation of a gain scheduled controller combined with a pattern recognition device to choose the most suitable gain.  While a very effective approach in some applications, the overall design is mainly verified by simulation.

The main goal of this paper is to present a type of learning control system for control affine nonlinear systems based on
a discretization of the {\em Chen-Fliess functional series} or {\em Fliess operator} \citep{Fliess_81,Fliess_83,Isidori_95}. It is well known that any
analytic control affine state space system in continuous-time has an input-output map with a Fliess operator representation.
Therefore, the structure of the proposed learning system contains learning units that are known {\em a priori} to be capable of approximating the input-output behavior of
the plant to an arbitrary desired accuracy \citep{Gray-et-al_NM17}. The learning system is also capable of incorporating a given physical model or can be used to
provide purely data-driven control. The approach is distinct from  model based adaptive control in that the plant is not made to track the
output of a reference model, and there is no adaptation of the model. It is also distinct from the local adaptive control approach in that there is no need to linearize
models or develop a gain scheduling strategy.
The method is demonstrated using a two-input, two-output Lotka-Volterra system. Some of these results have appeared in preliminary form
(often without proof) in \cite{Gray-etal_ICSTCC17,Gray-etal_CISS19,Gray-etal_ACC19,Venkatesh-etal_CDC2019}.

The paper is organized as follows. In the next section, a brief summary of the key concepts concerning discrete-time Fliess operators is given.
A type of learning unit is then described in Section~\ref{sec:LU} based on
discrete-time Fliess operators along with a purely inductive implementation.
In the subsequent section, it is shown how to combine this type of learning with model based control.
The main conclusions of the paper are given in the last section, as well as directions for future research.

\section{Discrete-Time Fliess Operators}
\label{sec:DTFO}

In this section, a brief review of discrete-time Fliess operators is presented.
For additional details, see \cite{Duffaut_Espinosa-etal_18,Gray-et-al_NM17}.

An {\em alphabet} $X=\{ x_0,x_1,$ $\ldots,x_m\}$ is any nonempty and finite set
of noncommuting symbols referred to as {\em
letters}. A {\em word} $\eta=x_{i_1}\cdots x_{i_k}$ is a finite sequence of letters from $X$.
The number of letters in a word $\eta$, written as $\abs{\eta}$, is called its {\em length}.
The empty word, $\emptyset$, is taken to have length zero.
The collection of all words having length $k$ is denoted by
$X^k$. Define $X^\ast=\bigcup_{k\geq 0} X^k$ and $X^{\leq J}=\bigcup_{k = 0}^J X^k$.
The former is a monoid under the concatenation product.
Any mapping $c:X^\ast\rightarrow
\re^\ell$ is called a {\em formal power series}.
Often $c$ is
written as the formal sum $c=\sum_{\eta\in X^\ast}(c,\eta)\eta$,
where the {\em coefficient} $(c,\eta)$ is the image of
$\eta\in X^\ast$ under $c$.
The set of all noncommutative formal power series over the alphabet $X$ is
denoted by $\allseriesell$. It forms an associative $\re$-algebra under the Cauchy product.

Inputs are assumed to be sequences of vectors from the normed linear space
\begdi
l_\infty^{m+1}(N_0):=\{\hat{u}=(\hat{u}(N_0),\hat{u}(N_0+1),\ldots): \norm{\hat{u}}_{\infty}<\infty\},
\enddi
where $\hat{u}(N):=[\hat{u}_0(N),\hat{u}_1(N),\ldots,\hat{u}_m(N)]^T$, $N\geq N_0$ with $\hat{u}_i(N)\in\re$,
$\abs{\hat{u}(N)}:=\max_{i\in\{0,1,\ldots,m\}}$ $\abs{\hat{u}_i(N)}$, and
$\norm{\hat{u}}_\infty:=\sup_{N\geq N_0}\abs{\hat{u}(N)}$.
The subspace of finite sequences over $[N_0,N_f]$ is denoted by $l_\infty^{m+1}[N_0,N_f]$.

\begde
Given a generating series $c\in\allseriesell$, the corresponding {\bf discrete-time Fliess operator} is defined as
\begdi 
\hat{F}_c[\hat{u}](N)=\sum_{\eta\in X^\ast} (c,\eta)S_{\eta}[\hat{u}](N)
\enddi
for any $N\geq N_0$, where
\begdi 
S_{x_i\eta}[\hat{u}](N)=\sum_{k=N_0}^N \hat{u}_{i}(k)S_{\eta}[\hat{u}](k)
\enddi
with $x_i\in X$,
$\eta\in X^\ast$, and
$\hat{u}\in l_\infty^{m+1}[N_0]$. By assumption,
$S_\emptyset[\hat{u}](N):=1$.
\endde

Following \cite{Grune-Kloeden_01}, select some fixed $u\in L_1^m[0,T]$ with $T>0$ finite.
Choose an integer $L\geq 1$, let $\Delta:=T/L$, and define the sequence of real numbers
\begdi 
\hat{u}_i(N)=\int_{(N-1)\Delta}^{N\Delta} u_i(t)\,dt,\;\;i=0,1,\ldots,m,
\enddi
where $N\in\{1,2,\ldots,L\}$.
Assume $u_0=1$ so that $\hat{u}_0(N)=\Delta$.
A truncated version
of $\hat{F}_c$ will be useful,
\begeq \label{eq:DT-truncated-Fliess-operator-defined}
\hat{y}(N)=\hat{F}^J_c[\hat{u}](N):=\sum_{\eta\in X^{\leq J}} (c,\eta) S_\eta[\hat{u}](N),
\endeq
since numerically only finite sums can be computed.
The main assertion proved in \cite[Theorems 6 and 7]{Gray-et-al_NM17} is that the class of truncated, discrete-time
Fliess operators acts as a set of universal approximators with computable error bounds for their continuous-time counterparts described
in \cite{Fliess_81,Fliess_83,Isidori_95}. In which case, they can be used to approximate any input-output system corresponding to an analytic control
affine state space realization
\begin{subequations} \label{eq:general-MIMO-control-affine-system}
\begin{align}
\dot{z}(t)&= g_0(z(t))+\sum_{i=1}^m g_i(z(t))\,u_i(t),\;\;z(t_0)=z_0 \label{eq:state} \\
y_j(t)&=h_j(z(t)),\;\;j=1,\ldots,\ell \label{eq:output}
\end{align}
\end{subequations}
with increasing accuracy as $L$ and $J$ increase.
This fact is
exploited in the next subsection to create a type of learning unit for data generated by such dynamical systems.

\section{Learning Unit Based on Discrete-Time Fliess Operator}
\label{sec:LU}

The main objective of this section is to introduce a learning unit based on a truncated discrete-time Fliess operator whose coefficients are identified
via a standard least-squares
algorithm. In general, there is one learning unit per output channel. So without loss of generality it is assumed that $\ell=1$.
First the basic architecture of the learning unit is described, and then
an inductive implementation of the underlying learning algorithm is developed.

\subsection{Learning Unit Architecture}

\begin{figure}[tb]
\begin{center}
\includegraphics[scale=0.37]{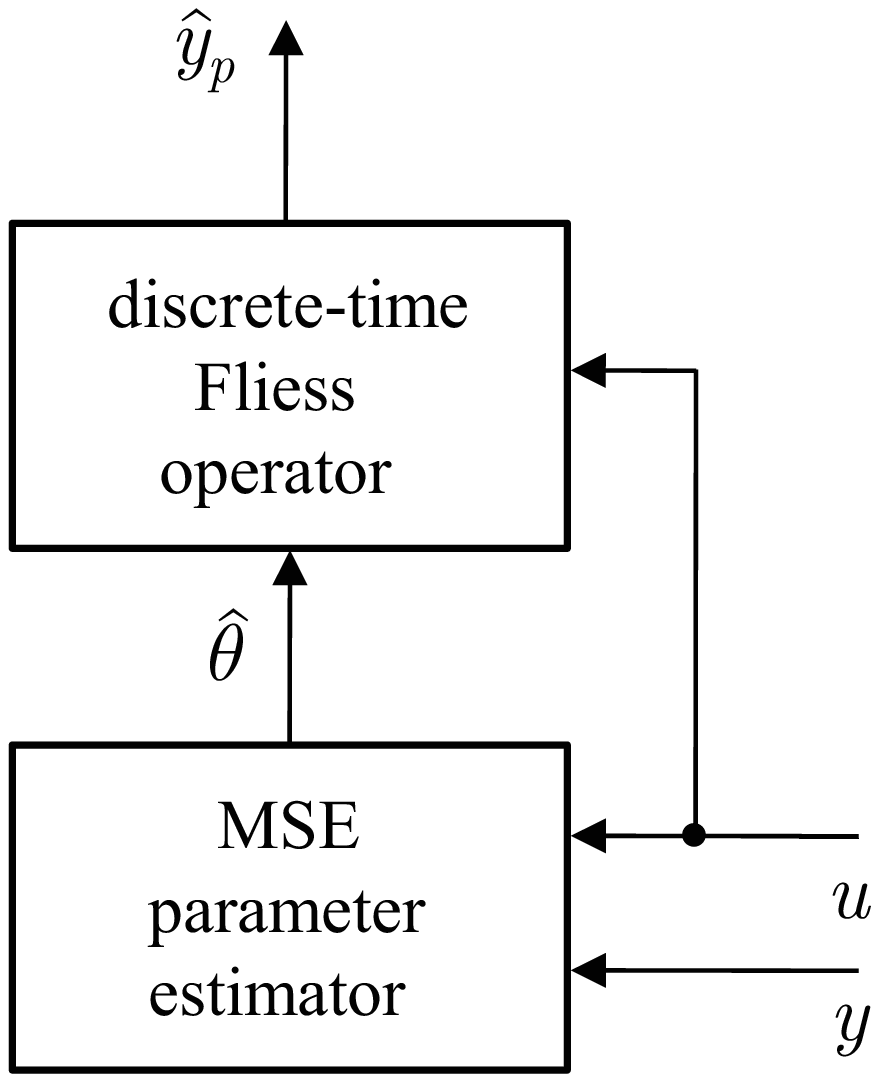}
\caption{Learning unit based on a discrete-time Fliess operator}
\label{fig:learning-unit}
\end{center}
\end{figure}

The first step is to write
\rref{eq:DT-truncated-Fliess-operator-defined} as an inner product
\begeq \label{eq:DT-FO_adaptive-form}
\hat{y}(N)=\phi^T(N)\theta_0,\;\;N\geq 1,
\endeq
where
\begin{align*}
\phi(N)&=[S_{\eta_1}[\hat{u}](N)\;S_{\eta_2}[\hat{u}](N)\cdots S_{\eta_l}[\hat{u}](N)]^T \\ 
\theta_0&=[(c,\eta_1)\;(c,\eta_2)\cdots (c,\eta_l)]^T
\end{align*}
with $l=\card(X^{\leq J})=\sum_{k=0}^J (m+1)^k=((m+1)^{J+1}-1)/m$ and assuming some ordering $(\eta_1,\eta_2,\ldots)$ has been imposed on the words in $X^\ast$.
If some estimate of $\theta_0$ is available at time $N-1$, say $\hat{\theta}(N-1)$, then \rref{eq:DT-FO_adaptive-form} gives a corresponding
estimate of $\hat{y}(N)$:
\begeq \label{eq:hatyp}
\hat{y}_p(N):=\phi^T(N)\hat{\theta}(N-1).
\endeq
The following least-squares algorithm is used to update the series coefficients:
\begin{subequations} \label{eq:MSE-estimator}
\begin{align}
\hat{\theta}(N)&=\hat{\theta}(N-1)+g(N-1)e(N) \\
e(N)&=y(N\Delta)-\phi^T(N)\hat{\theta}(N-1) \\
g(N-1)&=\frac{P(N-2)\phi(N)}{1+\phi^T(N)P(N-2)\phi(N)} \\
P(N-1)&=P(N-2)- \nonumber \\
&\hspace*{0.2in}\frac{P(N-2)\phi(N)\phi^T(N)P(N-2)}{1+\phi^T(N)P(N-2)\phi(N)}
\end{align}
\end{subequations}
for any $N\geq 1$ with the initial estimate $\hat{\theta}(0)$ given, and $P(-1)$ is any positive definite matrix $P_0$  \citep[p.~65]{Goodwin-Sin_09}.
Covariance resetting is done periodically to enhance convergence.
The corresponding learning unit is shown in Figure~\ref{fig:learning-unit}. Here input-output data $(u,y)$ from some unknown continuous-time plant
(or the error system between the plant and an assumed model) is fed into the unit. The only assumption is that the data came from a system
which has a Fliess operator representation, for example, any system modeled by \rref{eq:general-MIMO-control-affine-system}.
In general, the learning unit has no a priori knowledge of the system, so $\hat{\theta}(0)$ is initialized to
zero.
Setting $P_0=I$, it is known that this algorithm minimizes the performance index
\begdi
{\mathbf J}_{\bar{N}}(\theta):=\sum_{N=1}^{\bar{N}} [y(N\Delta)-\phi^T(N)\theta]^2+\frac{1}{2}\norm{\theta-\hat{\theta}(0)}^2
\enddi
with respect to the parameter $\theta$. It should be stated that since the model class consists of truncated versions of $\hat{F}_c$, there
is no reason to expect the parameter vector $\hat{\theta}(N)$ to converge to $c$ in any fashion as $N$ increases. But this is not a problem
since the only objective is to ensure that the underlying continuous-time
input-output map
$F_c$ is well approximated by $\hat{F}^J_{\hat{\theta}(N)}$. On the other hand, the approximation theory presented in \cite{Gray-et-al_NM17}  guarantees that
if the underlying system has such a Fliess operator representation then the true generating series $c$ is a feasible limit point for
the sequence $\hat{\theta}(N)$, $N\geq 0$.

\subsection{Inductive Implementation}

To devise an inductive implementation of the learning unit, it is necessary to identify the
algebraic structure underling the iterated sums in the definition of the discrete-time Fliess operator.
The starting point for this is the following concept.

\begde \cite{Gray-et-al_NM17}
Given any $N\geq N_0$ and $\hat{u}\in l_\infty^{m+1}(N_0)$, a {\bf discrete-time Chen series} is defined as
\begdi
S[\hat{u}](N,N_0)=\sum_{\eta\in X^\ast} \eta S_\eta[\hat{u}](N,N_0),
\enddi
where
\begeq \label{eq:iterated-sums}
S_{x_i\eta}[\hat{u}](N,N_0)=\sum_{k=N_0}^{N} \hat{u}_{i}(k)S_{\eta}[\hat{u}](k,N_0)
\endeq
with $x_i\in X$, $\eta\in X^\ast$, and $S_\emptyset[\hat{u}](N,N_0):=1$.
If $N_0=0$ then
$S[\hat{u}](N,0)$ is abbreviated as $S[\hat{u}](N)$.
\endde

Let $X$ be arbitrary and
define $\hat{u}_\eta(N)=\hat{u}_{i_k}(N)\cdots\hat{u}_{i_1}(N)$ for any $\eta=x_{i_k}\cdots x_{i_1}\in X^\ast$ and $N\geq N_0$ with $\hat{u}_\emptyset(N):=1$.
In addition, $c_u(N):=\sum_{\eta\in X^\ast} \hat{u}_\eta(N)\eta$.
Then
\begdi
S_{x_i\eta}[\hat{u}](N_0,N_0)=\hat{u}_{x_i}(N_0)S_{\eta}[\hat{u}](N_0,N_0)
\enddi
so that $S_\eta[\hat{u}](N_0,N_0)=\hat{u}_\eta(N_0)$, and thus, $S[\hat{u}](N_0,N_0)=c_u(N_0)$.

\begex \label{ex:one-letter-DT-Chen-series-part1}
If $X=\{x_1\}$ and $\hat{u}_{x_1}(N_0)=\hat{u}_1(N_0)$, then
\begdi
S[\hat{u}](N_0,N_0)=\sum_{k=0}^\infty (\hat{u}_1(N_0)x_1)^k=:(1-\hat{u}_1(N_0)x_1)^{-1}.
\enddi
\endex

A key observation is that a discrete-time Chen series $S[\hat{u}](N,N_0)$ satisfies a difference equation as described next
and proved in Appendix~\ref{app:Chen-monoid}

\begth \label{th:Su-update} \citep{Gray-etal_CISS19}
For any $\hat{u}\in l_\infty^{m+1}(N_0)$ and $N\geq N_0$
\begdi
S[\hat{u}](N+1,N_0)=c_u(N+1)S[\hat{u}](N,N_0)
\enddi
with $S[\hat{u}](N_0,N_0)=c_u(N_0)$ so that
\begeq
S[\hat{u}](N,N_0)=\overleftarrow{\prod_{i=N_0}^N} c_u(i), \label{eq:DT-Chen-series-directed-product}
\endeq
where $\overleftarrow{\prod}$ denotes a directed product from right to left.
\endth

\begex \label{ex:one-letter-DT-Chen-series-part2}
Consider the case in Example~\ref{ex:one-letter-DT-Chen-series-part1} where
$X=\{x_1\}$ and $\hat{u}_{x_1}(i)=\hat{u}_1(i)$ for all
$i\geq N_0$. Then
$c_u(i)=\sum_{k\geq 0} (\hat{u}_1(i) x_1)^k=(1-\hat{u}_1(i)x_1)^{-1}$ and
\begdi
S[\hat{u}](N,N_0)=(1-\hat{u}_1(N)x_1)^{-1}\cdots (1-\hat{u}_1(N_0)x_1)^{-1}.
\enddi
For instance,
\begin{align*}
S[\hat{u}](1,0)&=c_u(1)c_u(0) \\
&=1+(\hat{u}_1(1) + \hat{u}_1(0))x_1+(\hat{u}_1^2(1) + \\
&\hspace*{0.2in}\hat{u}_1(1) \hat{u}_1(0) + \hat{u}_1^2(0))x_1^2+ (\hat{u}_1^3(1) +  \\
&\hspace*{0.2in}\hat{u}_1^2(1)\hat{u}_1(0)+ \hat{u}_1(1) \hat{u}^2_1(0) + \hat{u}_1^3(0))x_1^3+\cdots
\end{align*}
In this case, $S[\hat{u}](N,N_0)$ is always a rational series \citep{Berstel-Reutenauer_88}.
\endex

\begin{figure}[tb]
\begin{center}
\includegraphics[scale=0.45]{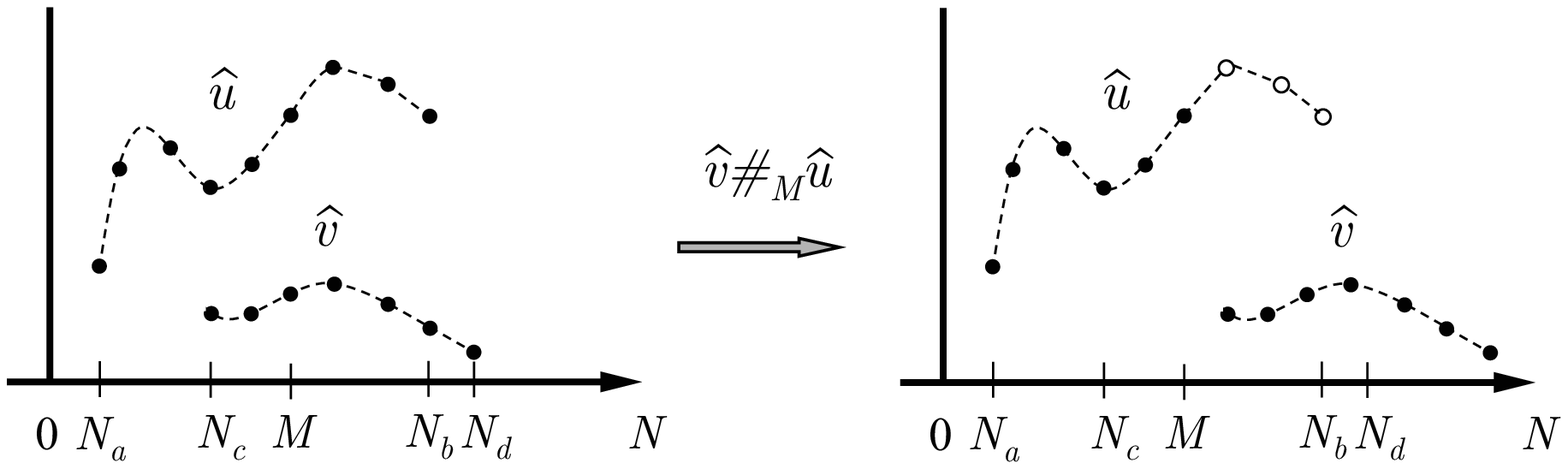}
\caption{Concatenation of $\hat{u}$ and $\hat{v}$}
\label{fig:sequence-catenation}
\end{center}
\end{figure}
Consider next two input sequences $(\hat{u},\hat{v})\in l_\infty^{m+1}[N_a,N_b]\times l_\infty^{m+1}[N_c,N_d]$ with $N_b>N_a$ and $N_d>N_c$.
The {\em concatenation} of $\hat{u}$ and $\hat{v}$
at $M\in[N_a,N_b]$ is taken to be
\begin{align*}
\lefteqn{(\hat{v}\#_{M}\hat{u})(N)} \\
&=\left\{\begar{rcl}
           \hat{u}(N) &:& N_a \leq N \leq M \\
           \hat{v}((N-M)+N_c) &:& M < N \leq M+(N_d-N_c)
         \endar\right.
\end{align*}
as shown in Figure~\ref{fig:sequence-catenation}.
Define the set of sequences
\begdi
l_{\infty,e}^{m+1}(0):=l_\infty^{m+1}(0)\cup \{\hat{\mathbf{0}}\},
\enddi
where $\hat{\mathbf{0}}$ denotes the empty sequence with duration zero so that formally
$\hat{v}\#_M\hat{\mathbf{0}}=\hat{\mathbf{0}}\#_M\hat{v}:=\hat{v}$
for all $\hat{v}\in l_{\infty,e}^{m+1}(0)$. In which case, $ l_{\infty,e}^{m+1}(0)$
is a monoid under this input concatenation operator. Define $S[\hat{\mathbf{0}}]=1$.
The following is a straightforward generalization of Theorem~\ref{th:Su-update}.

\begth \citep{Gray-etal_CISS19} (Discrete-time Chen's identity)
Given $(\hat{u},\hat{v})\in l_\infty^{m+1}[N_a,N_b]\times l_\infty^{m+1}[N_c,N_d]$,
$M\in [N_a,N_b]$, and $N\in[M,M+(N_d-N_c)]$ it follows that
\begdi
S[\hat{v}]((N-M)+N_c,N_c)S[\hat{u}](M,N_a)=S[\hat{v}\#_{M} \hat{u}](N,N_a).
\enddi
In particular, when $N_a=N_c=0$ then
\begeq
S[\hat{v}](N-M)S[\hat{u}](M)=S[\hat{v}\#_{M} \hat{u}](N). \label{eq:DT-Chen-identity}
\endeq
\endth

Define the set of discrete-time Chen series
\begin{align*}
{\cal M}_C(X)&=\{S[\hat{u}](N)\in\allseries: \hat{u}\in l_{\infty}^{m+1}[0,N_f],\\
&\hspace*{0.3in}0\leq N\leq N_f<\infty \}.
\end{align*}

\begth \citep{Gray-etal_CISS19}
${\cal M}_C(X)$ is a monoid under the Cauchy product.
In addition, $S:l^{m+1}_{\infty,e}(0)\rightarrow {\cal M}_C(X)$ is a monoid homomorphism.
\endth

\vspace*{-0.2in}

\begpr
The results follow directly from \rref{eq:DT-Chen-identity}.
\endpr

\vspace*{-0.2in}

Let ${\rm End}(\re^\infty)$ be the set of endomorphisms on the $\re$-vector space of real right-sided
infinite sequences. This set can be viewed as the monoid of doubly infinite matrices with well defined
matrix products and
unit $I=\diag(1,1,\ldots)$.
A monoid $M$ is said to have an {\em infinite dimensional real representation}, $\Pi$,
if the mapping $\Pi:M\rightarrow {\rm End}(\re^\infty)$ is a monoid homomorphism.
The representation
is {\em faithful} if $\Pi$ is injective.

\begth \label{th:MC-representation} \citep{Gray-etal_CISS19}
The monoid ${\cal M}_C(X)$ has a faithful infinite dimensional real representation $\Pi$ given by
$\Pi(S[\hat{u}](N))=\overleftarrow{\prod_{i=0}^N}{\cal S}(i)$,
where ${\cal S}(i)$ is any matrix representation of the $\re$-linear map on $\allseries$ given by
the left concatenation map
${\cal C}:d\mapsto c_u(i)d$.
\endth

\vspace*{-0.2in}

\begpr
The representation claim follows from \rref{eq:DT-Chen-series-directed-product}.
To see that $\Pi$ is injective, assume a fixed ordering of the words in $X^\ast$,
say $\{\eta_1,\eta_2,\ldots\}$. Then define the matrix
$[{\cal S}(i)]_{jk}=(c_u(i)\eta_k,\eta_j)=\hat{u}_\xi(i)$, where $\xi\eta_k=\eta_j$. Thus,
${\cal S}(i)$ is a lower triangular matrix with ones along the diagonal since $u_\emptyset(i)=1$, $i\geq 0$.
The first column is comprised of the coefficients of $c_u(i)$ in the order given to $X^\ast$. Hence, the
map $\Pi$ on the monoid ${\cal M}_C(X)$ is injective since $c_u(i)$ can be uniquely identified from ${\cal S}(i)=\Pi(S[\hat{u}](i,i))$.
\endpr

Note that the above theorem implies that \rref{eq:hatyp} can be written in the form
\begin{align}
\hat{y}_p(N+1)&=\hat{\theta}^T(N)\Pi(S[\hat{u}](N+1))e_1  \nonumber \\
&=\hat{\theta}^T(N){\cal S}(N+1)\Pi(S[\hat{u}](N))e_1 \label{eq:yp-Nplus1}
\end{align}
for $N\geq N_0$, where ${\cal S}(N+1)$ and $S[\hat{u}](N)$ have been suitable truncated, and $e_1:=[1\,0\,0\cdots 0]^T\in\re^l$.
Equation~\rref{eq:yp-Nplus1} can also be written in the form $\hat{y}_p(N+1)=Q(\hat{u}(N+1))$, where
$Q$ is a polynomial in the components of $\hat{u}(N+1)$ with maximum degree $l-1$.

\begex \label{ex:learning-example}
Suppose $X=\{x_1\}$ as in Example~\ref{ex:one-letter-DT-Chen-series-part2}.
Assuming the ordering on $X^\ast$ to be $\{\emptyset,x_1,x_1^2,\ldots\}$.
Then for all $i\geq 0$
\begdi
{\cal S}(i)=
\left[
\begin{array}{ccccc}
 1 & 0 & 0 & 0 & \cdots \\
 \hat{u}_1(i) & 1 & 0 & 0 & \cdots \\
 \hat{u}^2_1(i) & \hat{u}_1(i) & 1 & 0 & \cdots \\
 \hat{u}^3_1(i) & \hat{u}^2_1(i) & \hat{u}_1(i) & 1 & \cdots \\
 \vdots & \vdots & \vdots & \vdots & \ddots
\end{array}
\right]
\enddi
and $c_u(i)=\sum_{k\geq 0} \hat{u}^k_1(i)x_1^k$.
In addition,
\begin{align*}
\lefteqn{\Pi(S[\hat{u}](1))} \hspace*{0.2in}\\
&={\cal S}(1){\cal S}(0) \\
&=\left[
\begin{array}{c}
 1 \\
 \hat{u}_1(1)+\hat{u}_1(0) \\
 \hat{u}_1^2(1)+\hat{u}_1(1) \hat{u}_1(0)+\hat{u}_1^2(0) \\
  \hat{u}_1^3(1)+\hat{u}_1^2(1) \hat{u}_1(0)+\hat{u}_1(1) \hat{u}_1^2(0)+\hat{u}_1^3(0) \\
 \vdots \\
\end{array}
\right. \\
&\hspace*{-0.1in}\left.
\begin{array}{cccc}
0 & 0 & 0 & \cdots \\
1 & 0 & 0 & \cdots \\
 \hat{u}_1(1)+\hat{u}_1(0) & 1 & 0 & \cdots \\
 \hat{u}_1^2(1)+\hat{u}_1(1) \hat{u}_1(0)+\hat{u}_1^2(0) &  \hat{u}_1(1)+\hat{u}_1(0) & 1 & \cdots \\
 \vdots & \vdots & \vdots & \ddots
\end{array}
\right].
\end{align*}
As expected, the first column coincides with the coefficients of $S[\hat{u}](1)$ in Example~\ref{ex:one-letter-DT-Chen-series-part2}.
Setting $J=3$ so that $l=\card(X^{\leq J})=4$ gives the truncated versions
\begin{subequations} \label{eq:degree-3-DT-Chen-system}
\begin{align}
\hat{\theta}^T(N)&=\left[\!
\begin{tabular}{p{0.2in}p{0.27in}p{0.27in}p{0.27in}}
$(c,\emptyset)$ & $(c,x_1)$ & $(c,x_1^2)$ & $(c,x_1^3)$
\end{tabular}
\;\;\right] \\
{\cal S}(N)&=
\left[
\begin{array}{cccc}
 1 & 0 & 0 &0 \\
\hat{u}_1(N) & 1 & 0 & 0 \\
\hat{u}^2_1(N) & \hat{u}_1(N) & 1 & 0 \\
\hat{u}^3_1(N) & \hat{u}^2_1(N) & \hat{u}_1(N) & 1
\end{array}
\right] \\
\Pi(S[\hat{u}](N))&=
\left[
\begin{array}{cccc}
 1 & 0 & 0 & 0 \\
 S_{x_1}(N) & 1 & 0 & 0 \\
 S_{x_1^2}(N) & S_{x_1}(N) & 1 & 0  \\
 S_{x_1^3}(N) & S_{x_1^2}(N) & S_{x_1}(N) & 1
\end{array}
\right],
\end{align}
\end{subequations}
where $S_{x_1^k}(N):=(S[\hat{u}](N),x_1^k)$.
Therefore, the output
\begin{align*}
\hat{y}_p(N+1)&=Q(\hat{u}(N+1)) \\
&=\sum_{i=0}^3 q_i(N)\hat{u}_1^i(N+1),
\end{align*}
where the coefficients $q_i(N)$ are functions of $(c,x_1^k)$ and $S_{x_1^k}(N)$, $k=0,1,2,3$.

\begin{figure}[tb]
\vspace*{0.02in}
\begin{center}
\includegraphics[scale=0.55]{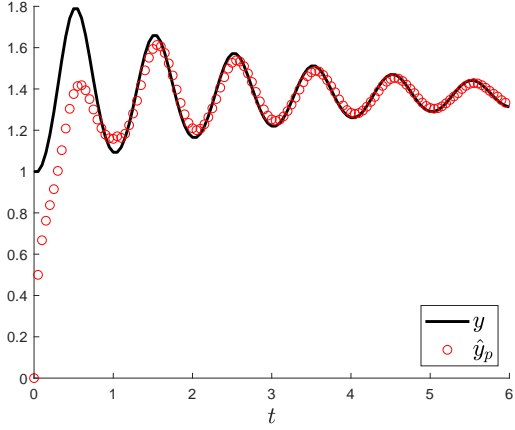}
\end{center}
\caption{Learning unit output $\hat{y}_p$ versus true output $y$ in Example~\ref{ex:learning-example}}
\label{fig:learning-example}
\end{figure}

As a specific example, consider a plant modeled by the Fliess operator
\begdi
y=F_c[u]=\sum_{k=0}^{\infty} (c,x_1^k)E_{x_1^k}[u](t,0),
\enddi
where the generating series is $c=\sum_{k\geq 0} x_1^k$, and
\begdi
E_{x_1^{k+1}}[u](t,0):=\int_0^t u(\tau)E_{x_1^{k}}[u](\tau,0)\:d\tau,\;\; k\geq 0
\enddi
with $E_{\emptyset}[u]:=1$.
The system has the state space realization
\begeq \label{eq:CT-rational-Ferfera-system-realization}
\dot{z}(t)=u(t),\;\; z(0)=0,\;\; y(t)=\expup^{z(t)}
\endeq
since for all $t\geq 0$
\begin{align*}
y(t)&=\sum_{k=0}^{\infty} E_{x_1}^k[u](t,0)\frac{1}{k!}=\sum_{k=0}^{\infty}\,E_{x_1^k}[u](t,0) = F_c[u](t).
\end{align*}
The output $y$ shown in Figure~\ref{fig:learning-example} is computed from a numerical simulation of the state
space model \rref{eq:CT-rational-Ferfera-system-realization}
when the input $u(t)=2e^{-t/3}\sin(2\pi t)$ is applied.
The output of the learning unit $\hat{y}_p(N)$, $N\geq 0$ as implemented using \rref{eq:MSE-estimator},
\rref{eq:yp-Nplus1}, and \rref{eq:degree-3-DT-Chen-system} is also shown in
the figure.  As the learning unit processes more data, its estimate of the output $y$ improves asymptotically.
\endex

The more challenging problem is systematically building a real representation of ${\cal M}_C(X)$ when $X$ has more than one letter, as in the multivariable case or
when the {\em drift letter} $x_0$ is present.
A partial ordering $\preceq$ is first defined on all words in \allwords. For all $\zeta,\eta\in \allwords$, let $\zeta \preceq \eta$ if and only if
there exists a $\gamma \in \allwords$ such that $\gamma^{-1}(\eta) = \zeta$, where $\gamma^{-1}$ denotes the left-shift operator. The following theorem is
proved in Appendix~\ref{app:poset}.

\begth \label{th:poset} \citep{Venkatesh-etal_CDC2019}
The pair $\left(\allwords,\preceq\right)$ is a partially ordered set.
\endth

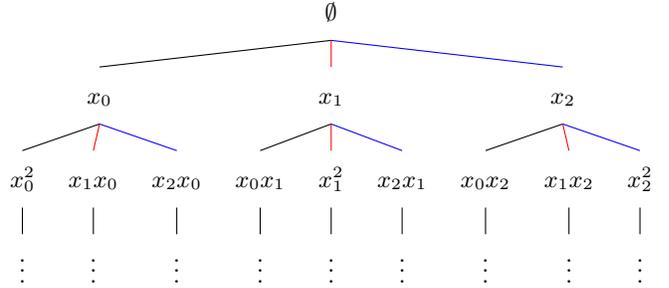
\begin{figure}[h!]
\vspace*{-0.1in}
\centering
{\small \begin{forest}
[$\emptyset$ [$x_0$ [$x_0^{2}$[$\vdots$]]
[$x_1x_0$, edge = {red}[$\vdots$]]
[$x_2x_0$, edge = {blue}[$\vdots$]]
  ]
[$x_1$, edge = {red} [$x_0x_1$[$\vdots$]]
[$x_1^{2}$, edge = {red} [$\vdots$]]
[$x_2x_1$, edge = {blue}[$\vdots$]]
  ]
 [$x_2$, edge = {blue} [$x_0x_2$[$\vdots$]]
[$x_1x_2$, edge = {red}[$\vdots$]]
[$x_2^{2}$, edge = {blue}[$\vdots$]]
  ]
  ]
\end{forest} }
\caption{Hasse diagram for $(\allwords,\preceq)$ when $X = \{x_0,x_1,x_2\}$}
\label{fig:Hasse-diagram}
\end{figure}

The partial order $(\allwords, \preceq)$ can be graphically represented by a Hasse diagram.  Starting with $\emptyset$ at the root, the Hasse diagram of $(\allwords, \preceq)$ when $ X = \{x_{0},x_{1},\hdots,x_{m}\}$ forms a ($m$+$1$)-ary infinitely branching tree. Define an injective map $R : X \longrightarrow {\cal C}$, where ${\cal C}$ is a set of colors. Color the edge between the nodes $\eta$ and $x_i\eta$ with the color $R(x_i)$ in the tree. As an illustration, the tree for the case when $m = 2$ is shown in Figure~\ref{fig:Hasse-diagram},  where $R(x_0) = \text{black}$, $R(x_1) = \text{red}$, and $R(x_2) = \text{blue}$.

In \rref{eq:yp-Nplus1}, the underlying discrete-time Fliess operator $\hat{F}_c\left[\hat{u}\right]$ has been truncated up to words of length $J$. Therefore, the tree is pruned at the $J$-th level. Next, a depth-first search (DFS) algorithm is employed to traverse the graph and generate words. The corresponding vector of words, $\chi^{J}(X)$, is called the \emph{order vector} of degree $J$ and is given by $\chi^0(X)=[\emptyset]$ and
\begin{equation}\label{eq:ord_vect_struct}
\chi^{J+1}(X)=
\begin{bmatrix}
\begin{array}{ccccc}
\emptyset & \chi^{J}(X) x_0 & \chi^{J}(X)  x_1 & \cdots &  \chi^{J}(X) x_m
\end{array}
\end{bmatrix}^{T}
\end{equation}
for $J\geq 0$.

\begex \label{ex:order_vec_2}
The tree for words $\eta \in X^{\leq 2}$ when $X = \{ x_0,x_1,x_2\}$ is given by
\begin{figure}[h!]
\vspace*{-0.15in}
\centering
{\small \begin{forest}
[$\emptyset$ [$x_0$ [$x_0^{2}$]
[$x_1x_0$, edge = {red}]
[$x_2x_0$, edge = {blue}]
  ]
[$x_1$, edge = {red} [$x_0x_1$]
[$x_1^{2}$, edge = {red}]
[$x_2x_1$, edge = {blue}]
  ]
 [$x_2$, edge = {blue} [$x_0x_2$]
[$x_1x_2$, edge = {red}]
[$x_2^{2}$, edge = {blue}]
  ]
    ]
\end{forest}}
\end{figure}

\newpage
\noindent
The DFS algorithm gives the order vector
\vspace*{-0.2in}
{\small
\begin{align*}
\lefteqn{\chi^{2}(X)}  \hspace*{0.1in}\\
&=\left[
\emptyset\;x_0\;x_0^2\;x_1x_0\;x_2x_0\;x_1\;x_0x_1\;x_1^2\;x_2x_1\;x_2\;x_0x_2\;x_1x_2\;x_2^2
\right]^T.
\end{align*}
}
\endex

Let ${\cal S}^J(N+1)$ denote the matrix ${\cal S}(N+1)$ truncated for words up to length $J$, i.e., ${\cal S}^J(N+1)\in\re^{l\times l}$ with
$l=\card(X^{\leq J})$.
An inductive algorithm to build such matrices is developed next.

\begde
Define $C_i$ as the colored tree of the Hasse diagram of $(\allwords,\preceq)$ up to the $i$-th level,
that is, the $(m+1)$-ary tree with $\emptyset$ as the root and $\eta \in X^i$ as leaves of the Hasse diagram.
Let $C := \{C_i : i \in \mathbb{N}_0\}$ be the set of colored trees given by $(\allwords,\preceq)$ of all levels.
\endde

It is useful to define a product $\dagger$ on $C$ as follows:
$C_i \dagger C_j := $\{tree with each leaf node $\beta \in X^i$ replaced by the tree $C_j$, where all the nodes of $C_j$
are right concatenated with $\beta$\}. The following theorem is proved in Appendix~\ref{app:dagger}.

\begth \label{th:dagger}
$(C,\dagger)$ is a commutative monoid isomorphic to the additive monoid $(\nat_0,+)$. Specifically,
$C_i \dagger C_j$ $ = C_{i+j}$ for all $C_i,C_j \in C$.
\endth

\begex \label{ex:tree_rec} Let $X = \{x_0,x_1\}$ and define the color map $R$ as: $R(x_0) = \text{red}$, $R(x_1) = \text{blue}$. Observe that

\begdi
C_1 =
\begin{minipage}{1in}
\begin{forest}
[$\emptyset$ [$x_0$, edge = {red} ]
[$x_1$, edge = {blue}]
 ]
\end{forest}
\end{minipage}
\enddi

\begdi
C_2 =
\begin{minipage}{1in}
\begin{forest}
[$\emptyset$ [$x_0$, edge = {red} [$x_0^{2}$, edge = {red}]
[$x_1x_0$, edge = {blue}]
  ]
[$x_1$, edge = {blue} [$x_0x_1$, edge = {red}]
[$x_1^{2}$, edge = {blue}]
  ]
    ]
\end{forest}
\end{minipage}
\enddi

\begdi
C_1 \dagger C_2 =
\begin{minipage}{1in}
\begin{forest}
[$\emptyset$ [$C_2x_0$, edge = {red} ]
[$C_2x_1$, edge = {blue}]
 ]
\end{forest}
\end{minipage}
\enddi

\noindent
The above tree can be expanded as

\begin{minipage}{1in}
\small
\begin{forest}
[$\emptyset$ [$x_0$, edge = {red} [$x_0^{2}$, edge = {red}[$x_0^{3}$, edge = {red}] [$x_1x_0^2$, edge = {blue}]]
[$x_1x_0$, edge = {blue} [$x_0x_1x_0$, edge = {red}] [$x_1^2x_0$, edge = {blue}]]
  ]
[$x_1$, edge = {blue} [$x_0x_1$, edge = {red}[$x_0^{2}x_1$, edge = {red}] [$x_1x_0x_1$, edge = {blue}]]
[$x_1^{2}$, edge = {blue} [$x_0x_1^{2}$, edge = {red}] [$x_1^3$,edge = {blue}]]
  ]
    ]
\end{forest} 
\end{minipage}
\\[0.1in]
This final tree is identified as $C_3$ so that $C_1 \dagger C_2 = C_3$.
\endex

Now assume that each color $R(x_i)$, $x_i\in X$, is given the weight $\hat{u}_i(N+1)$
at the discrete time instant $N+1$. Then it follows for any $\eta_j,\eta_k\in X^\ast$ that
\begin{align*}
[S(N+1)]_{jk}&= (c_u(N+1)\eta_k,\eta_j)  \\
&=\bigg\lbrace
\begin{array}{l}
  \text{weight of the path from $\eta_k$ to $\eta_j$ in $C_n$,}   \\
    \text{where} \hspace{0.2cm}  n\geq \vert\eta_j\vert.
\end{array}
\end{align*}
By Theorem \ref{th:dagger}, in the case where $X = \{x_0,x_1\}$ with color map $R$ defined as in
Example~\ref{ex:tree_rec}, $C_{J+1} = C_1 \dagger C_J$. That is,
\begdi
C_{J+1}=
\begin{minipage}{1in}
\begin{forest}
[$\emptyset$ [$C_Jx_0$, edge = {red} ]
[$C_Jx_1$, edge = {blue}]
 ]
\end{forest}

\end{minipage}
\enddi
Hence, from the structure of the order vector in \rref{eq:ord_vect_struct} and the above tree recursion,
one can deduce for any $m\geq 1$ that the block structure of the matrix ${\cal S}^{J+1}(N+1)$ can be written
inductively in terms of ${\cal S}^{J}(N+1)$ as:
\begin{align}
\lefteqn{{\cal S}^{J+1}(N+1)=} \nonumber \\
&\left[
   \begin{array}{c|c}
1                                       &       0 \cdots 0  \\ \hline
\hat{u}(N+1)\otimes({\cal S}^{J}(N+1)e_1) &
\begin{array}{c}
{\rm block\;diag}({\cal S}^{J}(N+1), \\
\quad \ldots,{\cal S}^{J}(N+1))
\end{array}
  \end{array}
\right], \label{eq:SN+1-matrix}
\end{align}
where `$\otimes$' denotes the Kronecker matrix product, and the block diagonal matrix is comprised on $m+1$ blocks.

\begex Let $X = \{x_0,x_1\}$. For $J=2$, the words are indexed by $\chi^{2}(X)=[\emptyset\;x_0\;x_0^2\;x_1x_0\;x_1\;x_0x_1\;x_1^2]$.
${\cal S}^{2}(N+1)$ can be computed directly from $C_2$ to be
\begin{equation} \label{eq:S2_N+1}
{\cal S}^{2}(N+1) =
\left[\begin{array}{c|ccccccc}
  1                  &  0         & 0 & 0 &        & 0         & 0 & 0 \\ \hline
  \hat{u}_0          &  1         & 0 & 0 & \vline & 0         & 0 & 0 \\
  \hat{u}_0^2        &  \hat{u}_0 & 1 & 0 & \vline & 0         & 0 & 0 \\
  \hat{u}_1\hat{u}_0 &  \hat{u}_1 & 0 & 1 & \vline & 0         & 0 & 0 \\ \cline{2-8}
  \hat{u}_1          &  0         & 0 & 0 & \vline & 1         & 0 & 0 \\
  \hat{u}_0\hat{u}_1 &  0         & 0 & 0 & \vline & \hat{u}_0 & 1 & 0 \\
  \hat{u}_1^2        &  0         & 0 & 0 & \vline & \hat{u}_1 & 0 & 1
  \end{array}\right].
\end{equation}
(For brevity, the argument $(N+1)$ is suppressed in the elements of the matrix.)
But the same matrix can also be computed inductively from \rref{eq:SN+1-matrix}.
For the base case, ${\cal S}^{0}(N+1) = 1$ so that
 \begin{equation*}
     {\cal S}^{1}(N+1) =
     \left[\begin{array}{ccccc}
                 1 & \vline & 0 &        & 0 \\ \hline
       \hat{u}_{0} & \vline & 1 & \vline & 0 \\ \cline{3-5}
       \hat{u}_{1} & \vline & 0 & \vline & 1
     \end{array}\right].
 \end{equation*}
Applying \rref{eq:SN+1-matrix} once more gives \rref{eq:S2_N+1}.
\endex

\section{Combining Learning and Model Based Control}
\label{sec:learning-control}

\begin{figure}[tb]
\begin{center}
\includegraphics[scale=0.3]{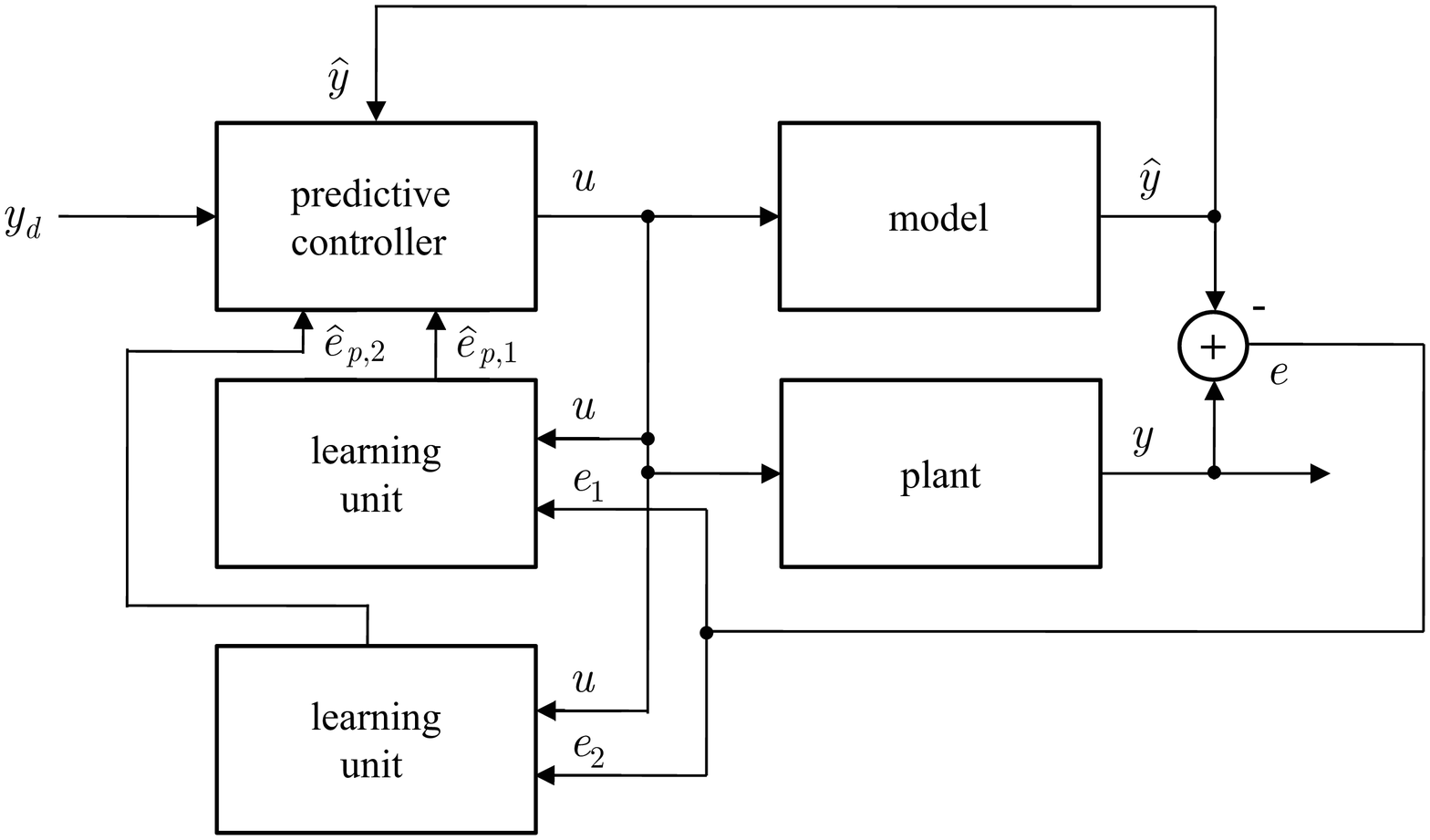}
\caption{Closed-loop system with a two-input, two-output predictive controller and two learning units}
\label{fig:MIMO-LU-control-system}
\end{center}
\end{figure}

Suppose $y_d$ is a desired output known to be in the range of a given plant with an underlying but unknown Fliess operator
representation $F_c$. It is most likely in applications that
$y_d$ was designed using an assumed model \rref{eq:general-MIMO-control-affine-system} with perhaps the aid of some expert knowledge.
Already this implies that the model is not too
poor an approximation of the plant, otherwise $y_d$ may not be in its range of the true plant. When both the plant and model are given the same input,
a modeling error $e_i=y_i-\hat{y}_i$ is generated for the $i$-th channel of the plant's output as shown in Figure~\ref{fig:MIMO-LU-control-system}.
This signal and the applied input
are then fed to a learning unit of the type
presented in the previous section in order to learn the input-output behavior of each error map $u\mapsto e_i$, $i=1,\ldots,m$,
which in this case must also have a Fliess operator representation.
At any time instant the output of the plant
is approximated by $\hat{y}+\hat{e}_p=\hat{y}+Q(\hat{u})$, where $Q$ was defined using \rref{eq:yp-Nplus1}.
A suitable input $u$ for tracking $y_d$ can be approximated by a piecewise constant
function taking values for $N\in\{1,2,\ldots,L\}$ equivalent to
\begeq \label{eq:preditive-controller}
\hat{u}(N):=\argminA_{\abs{\hat{u}(N)}\leq \bar{u}} y_e^T(N)Wy_e(N)
\endeq
for some fixed bound $\bar{u}>0$ and where
\begdi
y_e(N):=y_d(N\Delta)-[\hat{y}(N\Delta)+Q(\hat{u}(N))]
\enddi
with $W\in\re^{\ell\times\ell}$ being a fixed symmetric positive semi-definite weighting matrix.
The MatLab command {\tt fmincon} can be used to compute these local minima over the interval $[-\bar{u},\bar{u}]$.
In summary then equations \rref{eq:MSE-estimator}, \rref{eq:yp-Nplus1}, and \rref{eq:preditive-controller}
provide a fully inductive implementation of a one step ahead predictive controller with learning.
If the model is omitted from this set up, the resulting controller is still viable and can be viewed as a type of data-driven/model free closed-loop
system as first proposed for SISO systems in \cite{Gray-etal_ICSTCC17}.

As an example, consider the classical Lotka-Volterra model
\begeq
\dot{z}_i=\beta_i z_i+\sum_{j=1}^n \alpha_{ij}z_i z_j,\;\;i=1,\ldots,n,
\label{eq:LV1}
\endeq
used to describe
the population dynamics of
$n$ species in competition \citep{Chauvet-etal_02,May-Leonard_75,Smale_76}.
Here $z_i$ is the biomass of the $i$-th species, $\beta_i$ represents the growth rate of
the $i$-th species, and the parameter
$\alpha_{ij}$ describes the influence of the $j$-th species on the $i$-th species.
More recently in \cite{Jafarian-etal_18} it was shown that a power network,
where each node voltage $z_i$ is regulated by a quadratic droop controller, has
dynamics described by a Lotka-Volterra model.
In general, this model can exhibit
a wide range of behaviors including the presence of
multiple stable equilibria, stable limit cycles, and even chaotic behavior.

Consider the case where a subset of system parameters $\beta_{i_j}$, ${j=1,\ldots, m}$ in \rref{eq:LV1}
can be actuated and thus viewed as inputs $u_i$, $i=1,\ldots,m$.
Assume some set of output functions is given
\begeq
y_j=h_j(z),\;\;j=1,\ldots,\ell. \label{eq:LV4}
\endeq
Since the inputs enter the
dynamics linearly, it is clear that \rref{eq:LV1}-\rref{eq:LV4}
constitute a control affine analytic state space system.
In which case, the input-output map $u\mapsto y$
has an underlying Fliess operator representation $F_c$ with generating series
$c\in\allseriesell$ computable directly from \rref{eq:LV1}-\rref{eq:LV4} and a given initial condition $z_0$ \citep{Fliess_81,Fliess_83,Isidori_95}.
Of particular interest here is the special case of a predator-prey system, which is a two dimensional Lotka-Volterra system
\begin{align}
\dot{z}_1&=\beta_1 z_1-\alpha_{12}z_1z_2  \label{eq:LV-system_a} \\
\dot{z}_2&=-\beta_2 z_2+\alpha_{21}z_1z_2, \label{eq:LV-system_b}
\end{align}
where $y_1=z_1$ and $y_2=z_2$ are taken to be the population of prey and predator species respectively,
and \rref{eq:LV1} has been re-parameterized so that $\beta_i,\alpha_{ij}>0$.
This positive system has precisely
two equilibria when all the parameters are fixed, namely, a saddle point
equilibrium at the origin and a center at
$z_e=(\beta_2/\alpha_{21},\beta_1/\alpha_{12})$ corresponding to periodic solutions.

\begin{figure}[tb]
\begin{center}
\includegraphics[scale=0.55]{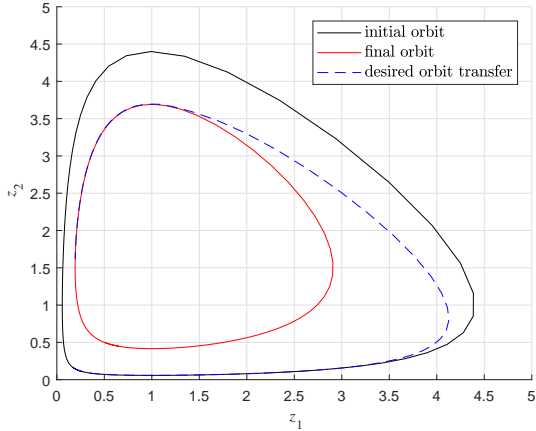}
\end{center}
\caption{Orbit transfer problem}
\label{fig:desired-orbit-transfer}
\end{figure}

\begin{table}[tb]
\begin{center}
Table 1.~Discretization parameters for simulations \\
\vspace*{0.1in}
\renewcommand{\arraystretch}{1.3}
\begin{tabular}{|c|c|c|c|c|}
\hline
$L$ & $J$ & $T$ & $\Delta$ & $\epsilon$  \\ \hline
100 & 3   & 6 & 0.06 & 0.05  \\ \hline
\end{tabular}
\end{center}
\end{table}

\begin{table}[tb]
\begin{center}
Table 2.~Normalized RMS tracking errors for MIMO system \\
\vspace*{0.1in}
\renewcommand{\arraystretch}{1.3}
\begin{tabular}[tb]{|c|c|c|c|c|}
\hline
$\Delta\alpha_{12}$& $\Delta\alpha_{21}$& $\delta y_1$  & $\delta y_2$ & $\|\hat{u}\|_{\infty}$  \\
\hline\hline
\multicolumn{2}{|c|}{exact model} & 8.66$\times10^{-9}$ & 1.25$\times10^{-8}$ & 2\\
\hline
-5 & 0 & 0.012 & 0.007 & 2  \\
\hline
0 & -5 & 0.020 & 0.016 & 2 \\
\hline
5 & 0 & 0.004 & 0.006 & 2 \\
\hline
0 & 5 & 0.018 & 0.015 & 2\\
\hline
-10 & 0 & 0.016 & 0.012 & 1.5 \\
\hline
0 & -10 & 0.056 & 0.041 & 1.5 \\
\hline
10 & 0 & 0.010 & 0.009 & 1.5 \\
\hline
0 & 10 & 0.037 & 0.025 & 1.5 \\
\hline
-20 & 0 & 0.023 & 0.024 & 0.5\\
\hline
 0 & -20 & 0.144 & 0.113 & 0.5 \\
 \hline
 20 & 0 & 0.012 & 0.016 & 0.5 \\
\hline
0 & 20 & 0.071 & 0.047 & 0.5 \\
\hline
-50 & 0 & 0.092 & 0.096 & 0.5\\
\hline
50 & 0 & 0.010 & 0.028 & 0.5 \\
\hline
0 & 50 & 0.062 & 0.095 & 0.5 \\
\hline
\multicolumn{2}{|c|}{model free} & 0.191 & 0.897 & 1 \\ \hline
\end{tabular}
\end{center}
\end{table}

\begin{figure}[tb]
\begin{center}
\includegraphics[scale=0.55]{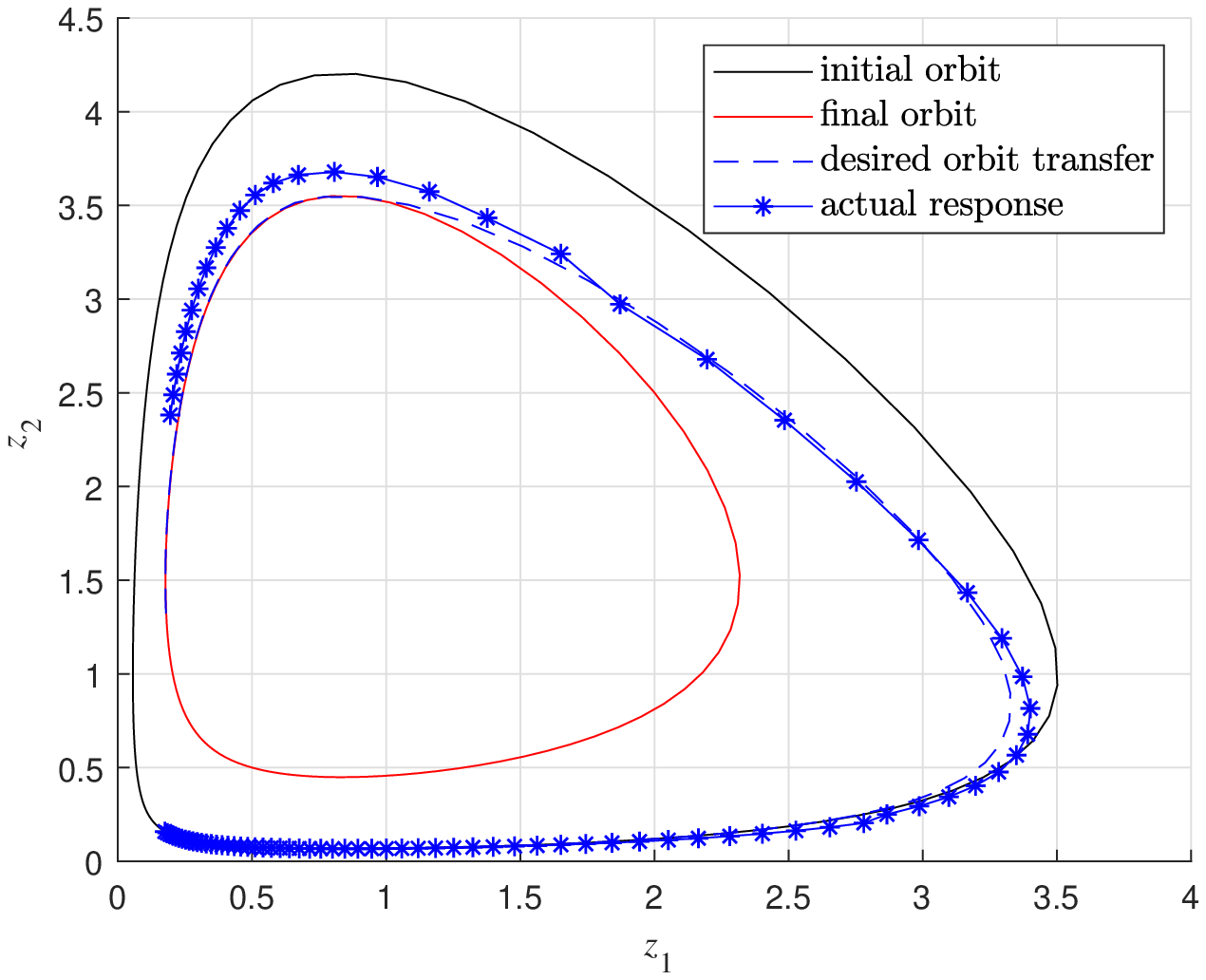}
\end{center}
\caption{Orbit transfer with +20\% model error in $\alpha_{21}$}
\label{fig:case1-orbit-transfer}
\end{figure}

\begin{figure}[tb]
\begin{center}
\includegraphics[scale=0.55]{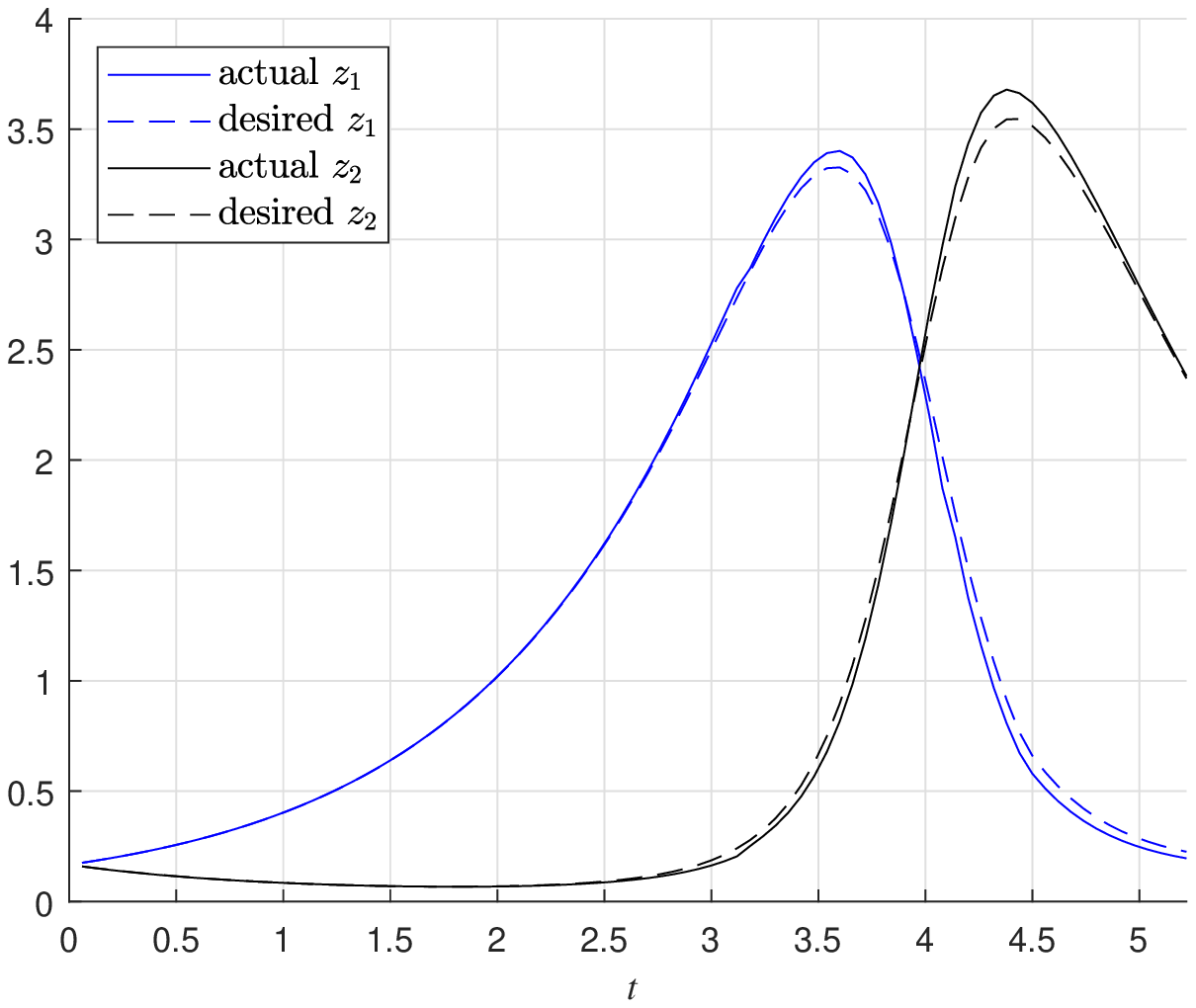}
\end{center}
\caption{State trajectories with +20\% model error in $\alpha_{21}$}
\label{fig:case1-state-traject}
\end{figure}

\begin{figure}[tb]
\vspace*{0.05in}
\begin{center}
\includegraphics[scale=0.55]{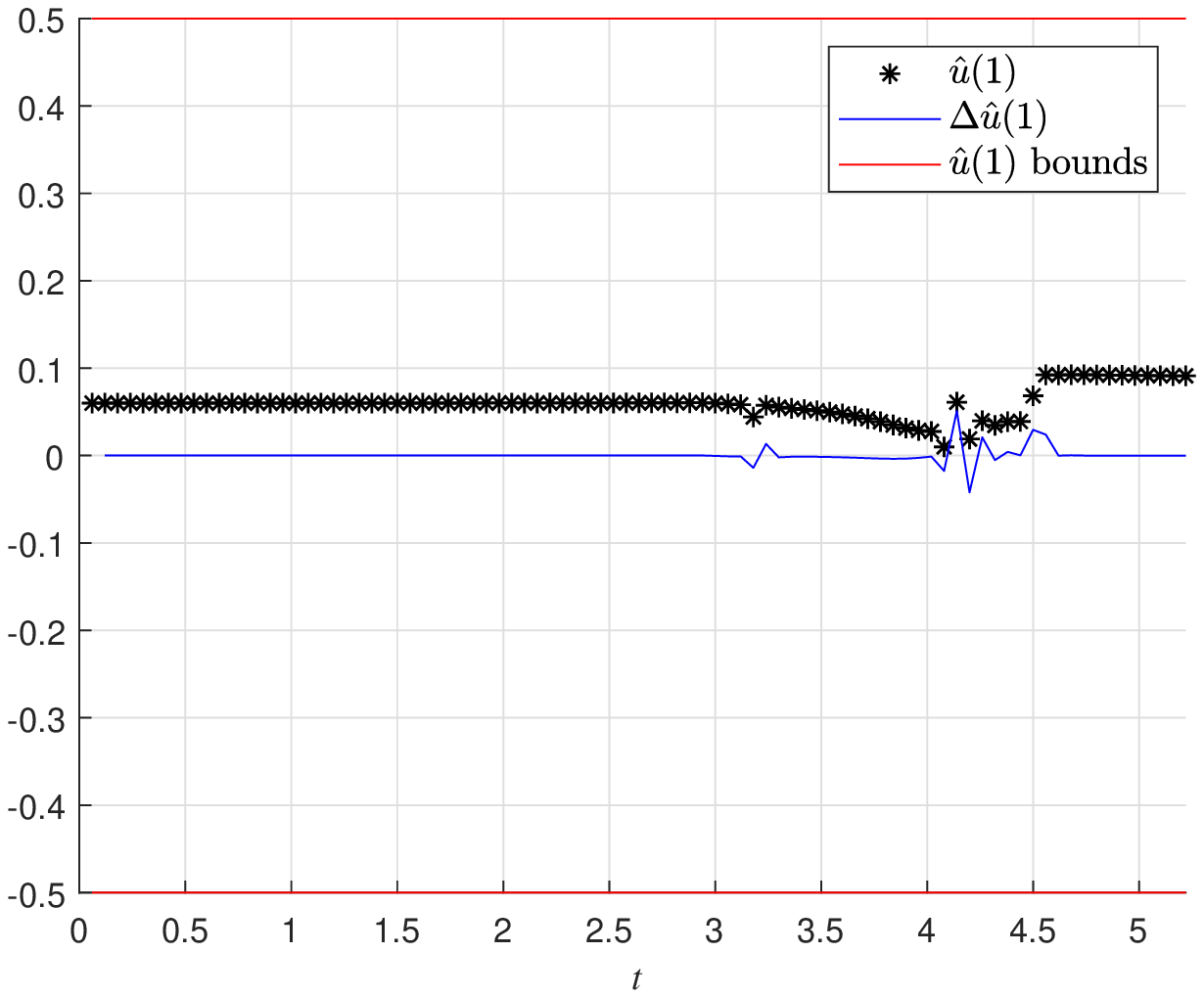} \\[0.05in]
\includegraphics[scale=0.55]{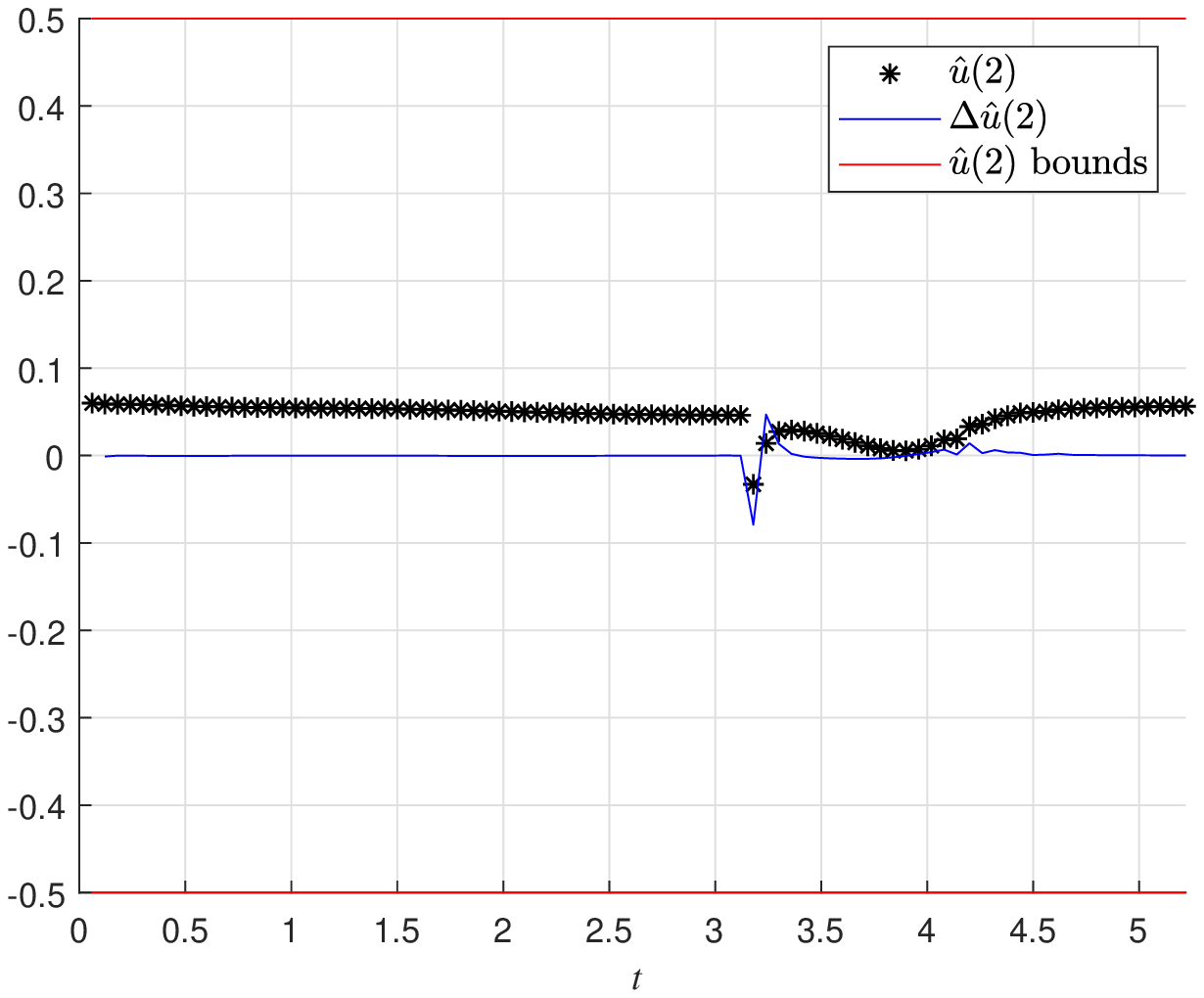}
\end{center}
\caption{Applied input with +20\% model error in $\alpha_{21}$}
\label{fig:case1-inputs}
\end{figure}

\begin{figure}[tb]
\vspace*{0.1in}
\begin{center}
\includegraphics[scale=0.55]{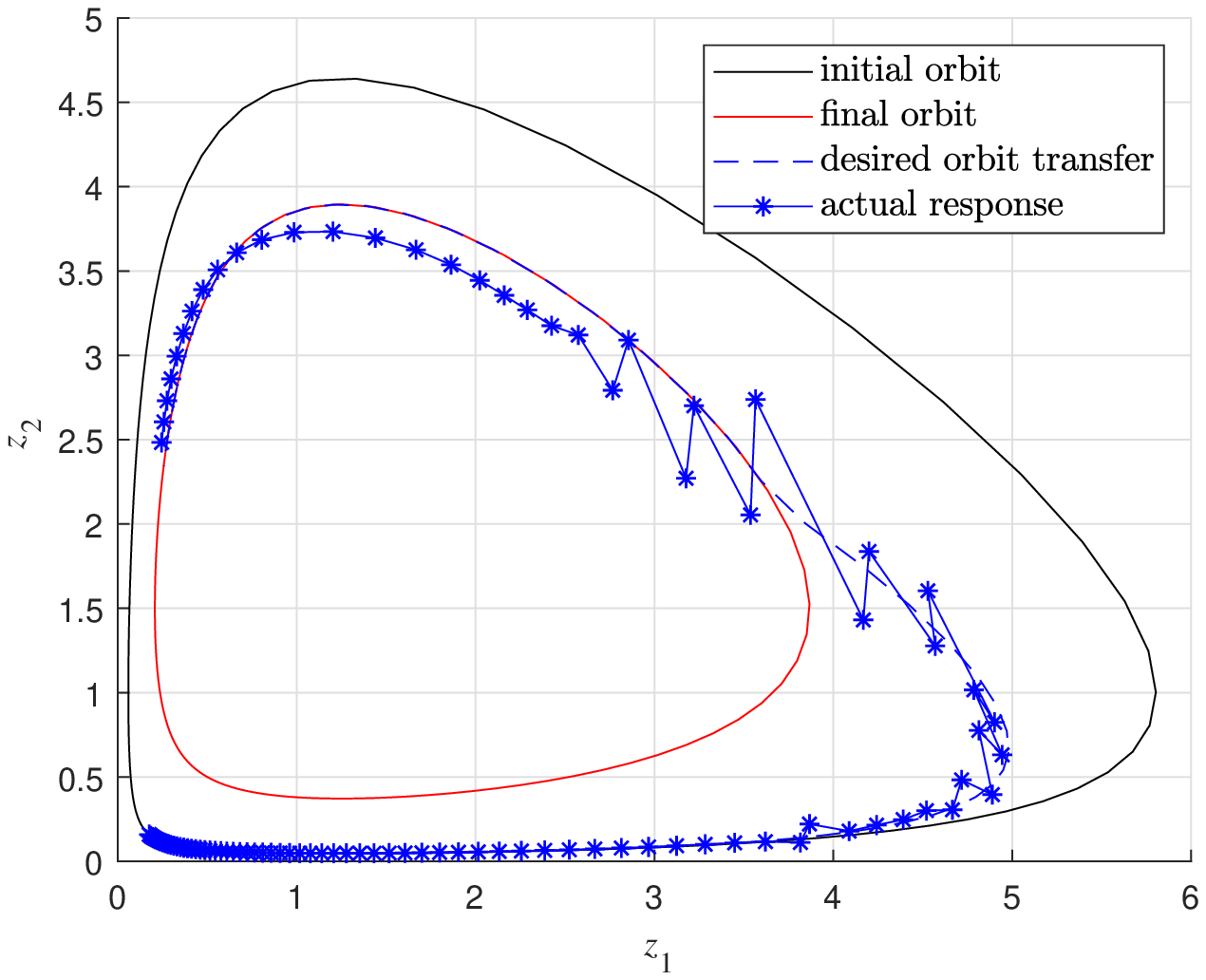}
\end{center}
\caption{Orbit transfer with -20\% model error in $\alpha_{21}$}
\label{fig:case3-orbit-transfer}
\end{figure}

\begin{figure}[tb]
\vspace*{0.05in}
\begin{center}
\includegraphics[scale=0.55]{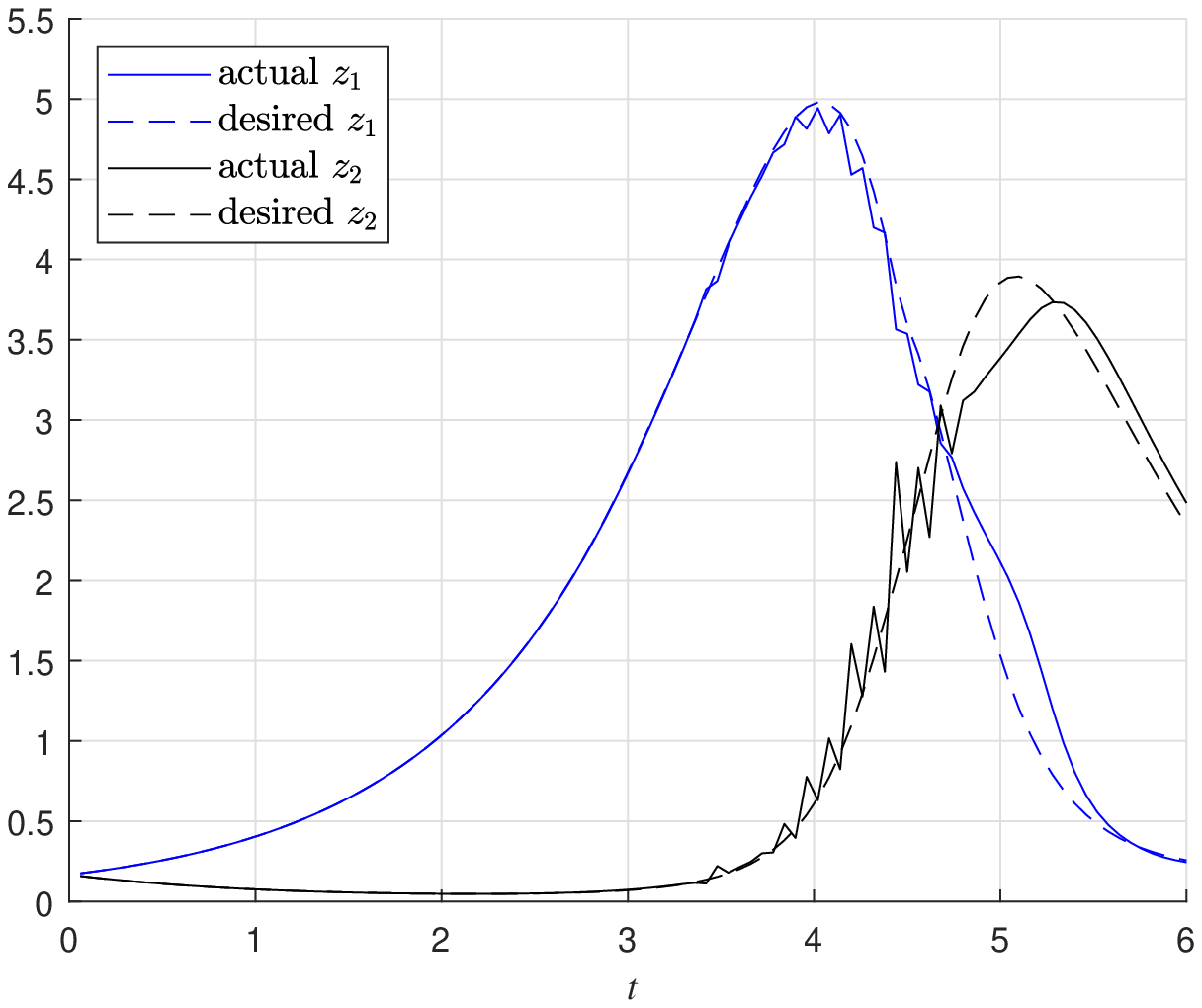}
\end{center}
\caption{State trajectories with -20\% model error in $\alpha_{21}$}
\label{fig:case3-state-traject}
\end{figure}

\begin{figure}[tb]
\begin{center}
\includegraphics[scale=0.55]{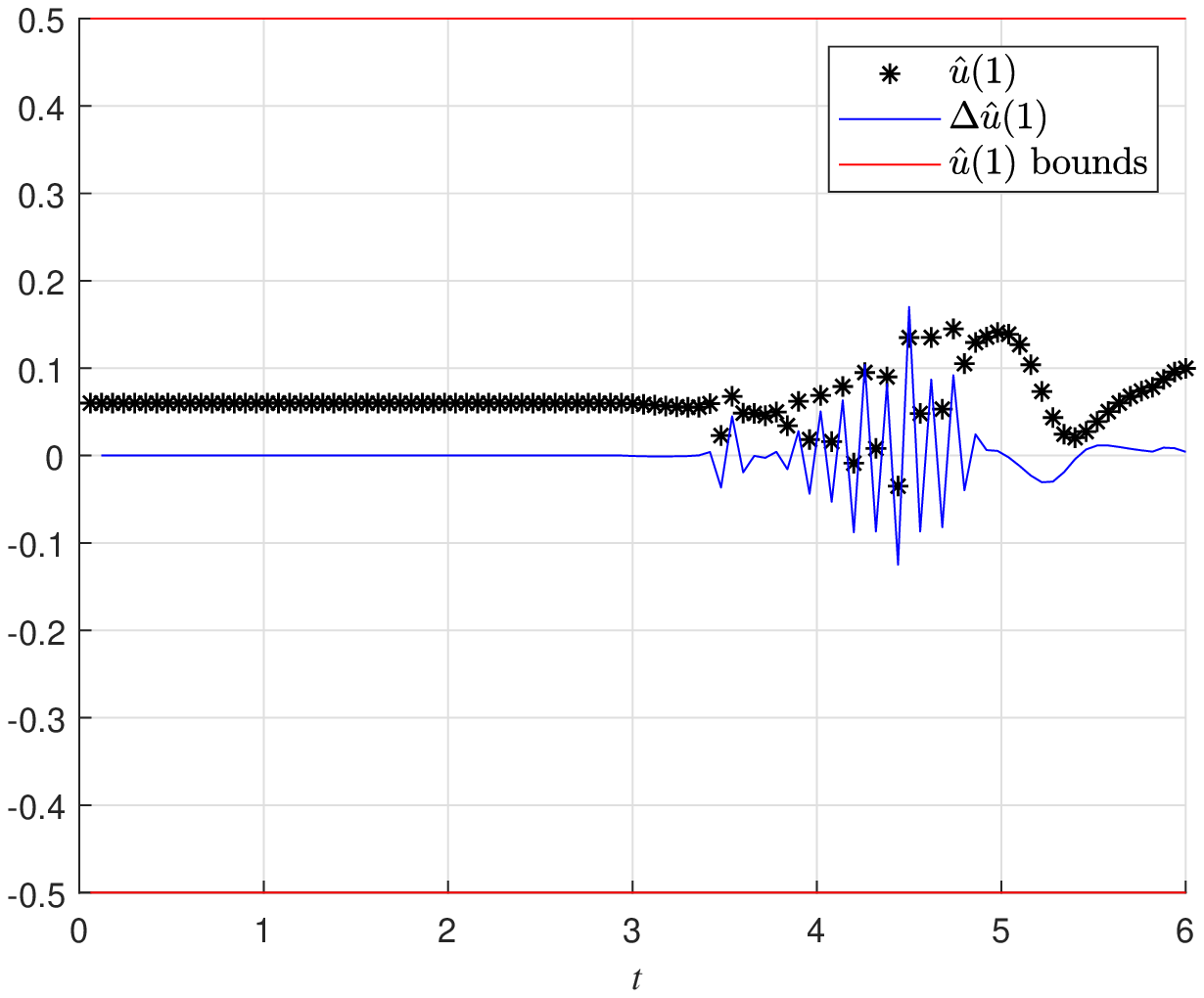} \\[0.05in]
\includegraphics[scale=0.55]{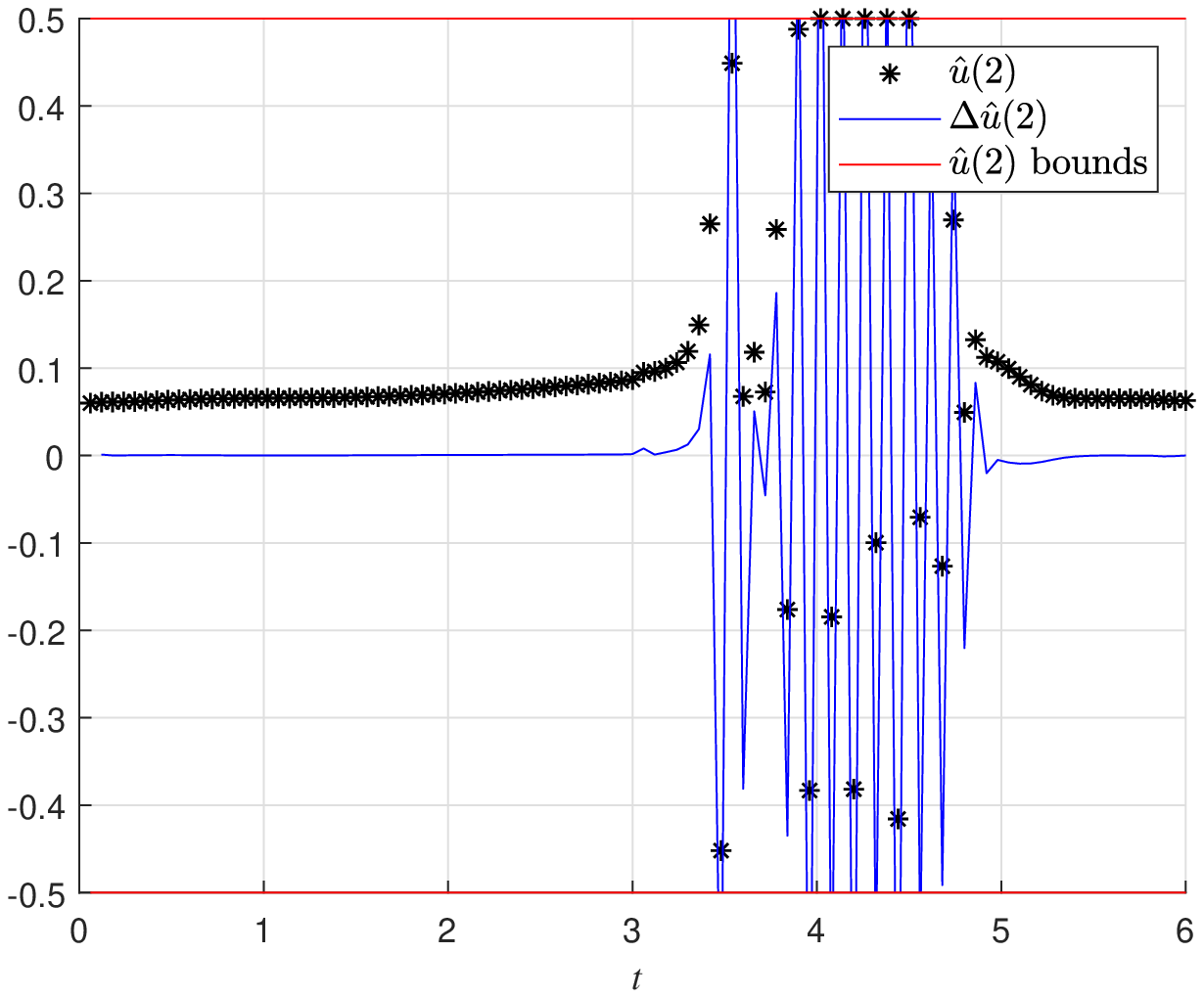}
\end{center}
\caption{Applied input with -20\% model error in $\alpha_{21}$}
\label{fig:case3-inputs}
\end{figure}

\begin{figure}[tb]
\vspace*{0.05in}
\begin{center}
\includegraphics[scale=0.55]{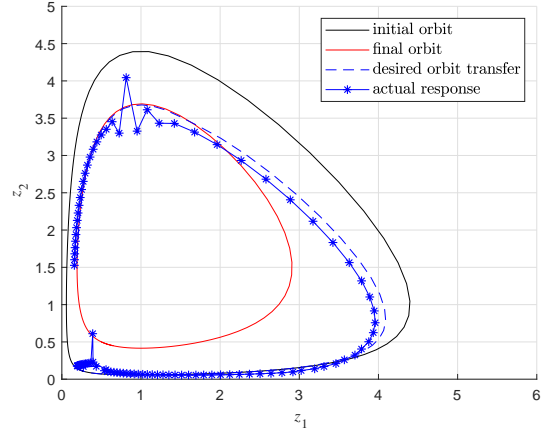}
\end{center}
\caption{Orbit transfer with no model}
\label{fig:model-free-orbit-transfer}
\end{figure}

\begin{figure}[tb]
\vspace*{0.1in}
\begin{center}
\includegraphics[scale=0.55]{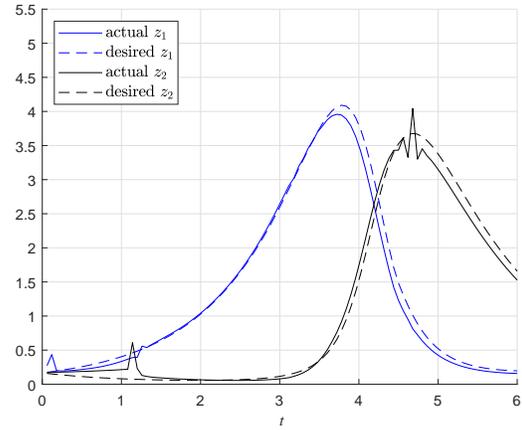}
\end{center}
\caption{State trajectories with no model}
\label{fig:model-free-state-traject}
\end{figure}

\begin{figure}[tb]
\begin{center}
\includegraphics[scale=0.55]{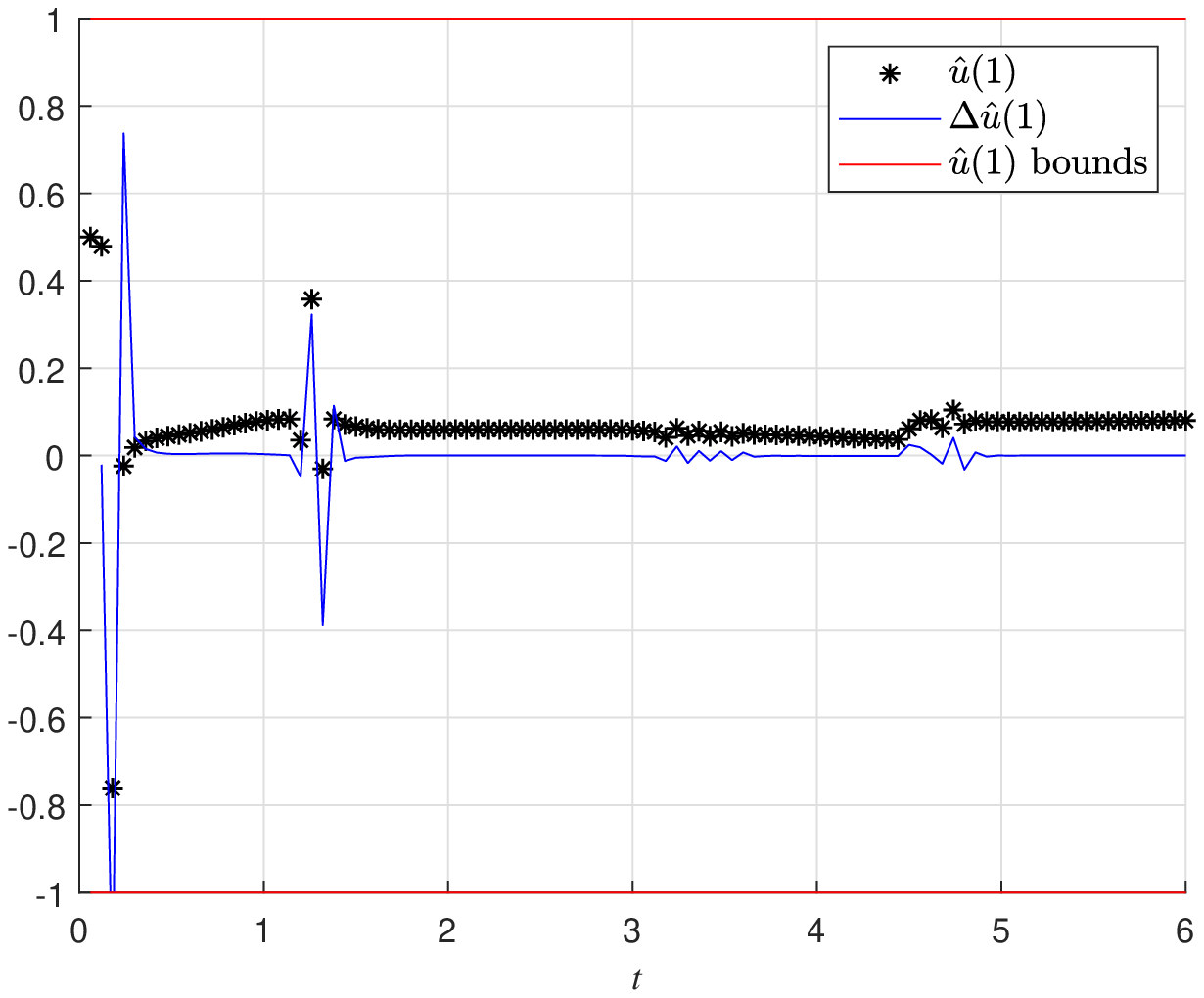} \\[0.05in]
\includegraphics[scale=0.55]{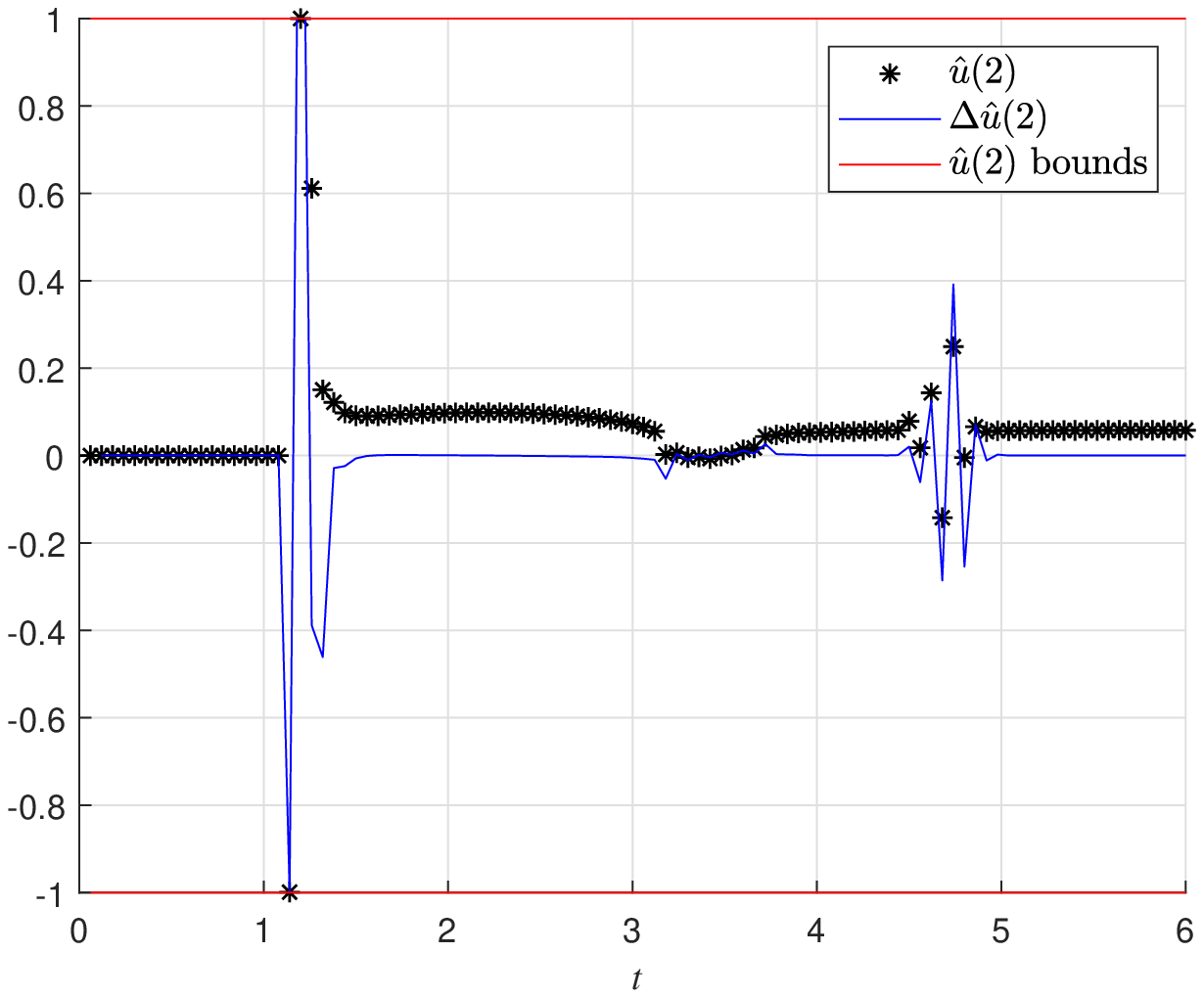}
\end{center}
\caption{Applied input with no model}
\label{fig:model-free-inputs}
\end{figure}

Taking the system inputs in \rref{eq:LV-system_a} and  \rref{eq:LV-system_b} to be $u_1=\beta_1$ and $u_2=\beta_2$,
the \emph{orbit transfer} problem as shown in Figure~\ref{fig:desired-orbit-transfer} is to determine an input
to drive the system from some initial orbit to within an $\epsilon$ neighborhood of a final orbit using a given orbit transfer trajectory.
The proposed controller was tested in simulation assuming all the
plant's parameters are set to unity. The discretization parameters were selected as in
Table~1, and the sampled input was bounded by $\|\hat{u}\|_\infty$ as given in Table~2 (the positivity constraint on the input was not enforced). The tracking performance for various choices of model
parameter errors is shown in Table~2.
For each plant parameter $\lambda$, $\Delta\lambda := (\lambda_{\text{model}} - \lambda_{\text{plant}})\times100\%$. In addition, $\delta y_{i}$ for $i = 1,2$ is the
RMS error per sample normalized by the sample value of desired trajectory for the given output channel. First the control system was tested assuming the exact plant model is available. In which case,
the modeling error is zero and the learning units are inactive. The closed-loop performance is therefore determined solely by the predictive controller, which is quite
accurate as shown in Table~2. Next a variety of parametric errors were introduced in the model. For all such cases, the weighting matrix $W$ was set to the identity matrix.
As an example, the simulation results for the case of $+20\%$ error in $\alpha_{21}$ are shown in
Figures~\ref{fig:case1-orbit-transfer}-\ref{fig:case1-inputs}.
The performance of the system for the case where there is $-20\%$ error in $\alpha_{12}$ was very similar.
The case where $\Delta\alpha_{21} = -20\%$ is an extreme scenario as decreasing $\alpha_{21}$ any further resulted in the plant's response being oscillatory. The simulation results pertaining to this case are shown in Figures~\ref{fig:case3-orbit-transfer}-\ref{fig:case3-inputs}.
Finally, the model free case was also simulated as shown in Figures~\ref{fig:model-free-orbit-transfer}-\ref{fig:model-free-inputs}.
Here it was necessary to select a nontrivial weighting matrix, in this case $W=[1\; 0.25;\;0.25\;1]$, in order for the optimizer to compensate for the cross coupling between the input-output channels, something that was done automatically when a model was present.
Note that
tracking was achieved, but the performance was about an order of magnitude worse
than most cases employing a model. While in practice input bounds are dictated by the physical application, it was
observed here that the larger the modeling error, the more conservative the input bounds needed to be in order to avoid
instabilities. On the other hand, if the bounds were too
conservative, then there was not enough actuation energy available to follow the desired trajectory.
For the sake of comparison to earlier work reported in \cite{Gray-etal_ACC19}, where only a single learning unit was employed
and thus only SISO and SIMO control (with $W = [1\;0.25;\;0.25\; 2]$) was possible, the simulation results are summarized in Tables~3-5. As a general rule, the
lack of a second control input rapidly reduced
performance when parameter errors exceeded five percent. But of course controller complexity was also significantly reduced in this case.


\begin{table}[b!]
\begin{center}
Table 3.~Normalized RMS tracking errors for SISO system $u_1\mapsto y_1$ \\
\vspace*{0.1in}
\renewcommand{\arraystretch}{1.3}
\begin{tabular}[tb]{|c|c|c|c|c|c|}
\hline
$\Delta\alpha_{12}$ & $\Delta\alpha_{21}$ & $\Delta\beta_2$ & $\delta y_1$ & $\delta y_2$ & $\|\hat{u}\|_{\infty}$  \\
\hline\hline
\multicolumn{3}{|c|}{exact model} & 1.547$\times 10^{-5}$ & 1.165$\times 10^{-4}$ & 1.4  \\ \hline
-5 & 0 & 0                        & 0.071 & 0.003                                 & 1.4  \\ \hline
0 & -5 & 0                        & 0.024 & 0.118                                 & 1.4  \\ \hline
0 & 0 & -5                        & 0.010 & 0.157                                 & 1.4  \\ \hline
\multicolumn{3}{|c|}{model free}  & 0.602 & 0.330                                 & 2  \\ \hline
\end{tabular}
\end{center}
\end{table}

\begin{table}[b!]
\begin{center}
Table 4.~Normalized RMS tracking errors for SISO system $u_1\mapsto y_2$ \\
\vspace*{0.1in}
\renewcommand{\arraystretch}{1.3}
\begin{tabular}[tb]{|c|c|c|c|c|c|}
\hline
$\Delta\alpha_{12}$ & $\Delta\alpha_{21}$ & $\Delta\beta_2$ & $\delta y_1$ & $\delta y_2$ & $\|\hat{u}\|_{\infty}$  \\
\hline\hline
\multicolumn{3}{|c|}{exact model} & 0.007 & 1.93$\times 10^{-7}$ & 1.4  \\ \hline
-5 & 0 & 0                        & 0.208 & 0.005                & 1.4  \\ \hline
0 & -5 & 0                        & 0.162 & 0.016                & 1.4  \\ \hline
0 & 0 & -5                        & 0.544 & 0.010                & 1.4  \\ \hline
\multicolumn{3}{|c|}{model free}  & 1.970 & 1.680                & 2  \\ \hline
\end{tabular}
\end{center}
\end{table}

\begin{table}[tb]
\begin{center}
Table 5.~Normalized RMS tracking errors for SIMO system $u_1\mapsto y$ \\
\vspace*{0.1in}
\renewcommand{\arraystretch}{1.3}
\begin{tabular}[tb]{|c|c|c|c|c|c|}
\hline
$\Delta\alpha_{12}$ & $\Delta\alpha_{21}$ & $\Delta\beta_2$ & $\delta y_1$ & $\delta y_2$ & $\|\hat{u}\|_{\infty}$  \\
\hline\hline
\multicolumn{3}{|c|}{exact model} & 8.73$\times 10^{-5}$ & 1.599$\times 10^{-4}$ & 2  \\ \hline
-5 & 0 & 0                        & 0.009 & 0.002                    & 1.4  \\ \hline
0 & -5 & 0                        & 0.094 & 0.118                    & 1.4  \\ \hline
0 & 0 & -5                        & 0.071 & 0.089                    & 1.4  \\ \hline
\multicolumn{3}{|c|}{model free}  & 0.167    & 0.055               & 1  \\ \hline
\end{tabular}
\end{center}
\end{table}

\section{Conclusions and Future Work}

A learning system for nonlinear control was presented based on discrete-time Chen-Fliess series and capable of
incorporating a given physical model. A fully inductive implementation
in the multivariable case required one to exploit the
underlying noncommutative algebraic and combinatorial structures in order to identify a
convenient basis to represent the learning dynamics. The method was demonstrated using a two-input, two-output Lotka-Volterra system.

Future work will include the introduction of measurement noise in the system, exercising the method on more complex engineering plants,
and identifying conditions under which closed-loop stability can be guaranteed.

\begin{ack}
This research was supported by the National Science
Foundation under grants CMMI-1839378 and CMMI-1839387.
\end{ack}

\appendix

\section{Proof of Theorem~\ref{th:Su-update}}
\label{app:Chen-monoid}

The first identity is addressed by proving that
\begdi
S_\eta[\hat{u}](N+1,N_0)=(c_u(N+1)S[\hat{u}](N,N_0),\eta),\;\;\forall \eta\in X^\ast
\enddi
via induction on the length of $\eta$.
When $\eta=\emptyset$ then trivially
$S_{\emptyset}[\hat{u}](N+1,N_0)=1=\hat{u}_\emptyset(N+1)S_\emptyset[\hat{u}](N,N_0)$.
If $\eta=x_i\in X$ then from \rref{eq:iterated-sums}
\begin{align*}
S_{x_i}[\hat{u}](N+1,N_0)&=\hat{u}_{x_i}(N+1)+S_{x_i}[\hat{u}](N,N_0) \\
&=\sum_{x_i=\xi\nu}\hat{u}_\xi(N+1) S_\nu[\hat{u}](N,N_0) \\
&=(c_u(N+1)S[\hat{u}](N,N_0),x_i).
\end{align*}
Finally, assume the identity holds for all words up to some fixed length $n\geq 0$. Then for
any $\eta\in X^n$ and $x_i\in X$ it follows that
\begin{align*}
\lefteqn{S_{x_i\eta}[\hat{u}](N+1,N_0)} \hspace*{0.3in}\\
&=\hat{u}_{x_i}(N+1)S_{\eta}[\hat{u}](N+1,N_0)+S_{x_i\eta}[\hat{u}](N,N_0) \\
&=\sum_{\eta=\xi\nu}\hat{u}_{x_i}(N+1)\hat{u}_\xi(N+1) S_\nu[\hat{u}](N,N_0)+ \\
&\hspace*{0.2in}\hat{u}_\emptyset(N+1) S_{x_i\eta}[\hat{u}](N,N_0)\\
&=\sum_{x_i\eta=\xi\nu}\hat{u}_\xi(N+1) S_\nu[\hat{u}](N,N_0) \\
&=(c_u(N+1)S[\hat{u}](N,N_0),x_i\eta),
\end{align*}
which proves the claim for all $\eta\in X^\ast$. The second identity in the theorem follows directly from the first.

\section{Proof of Theorem~\ref{th:poset}}
\label{app:poset}

Let $\eta,\zeta,\gamma,\alpha,\beta \in \allwords$. Reflexivity is trivial since
$\emptyset^{-1}(\eta) = \eta$ if and only if $\eta \preceq \eta$.
To prove transitivity, first observe that
\begin{align*}
      (\eta \preceq \zeta) &\Rightarrow \exists \beta : \beta^{-1}(\zeta) = \eta   \\
      (\zeta \preceq \gamma) &\Rightarrow \exists \alpha : \alpha^{-1}(\gamma) = \zeta
\end{align*}
so that
\begdi
((\alpha\beta)^{-1}(\gamma) = \eta) \Rightarrow \eta \preceq \gamma.
\enddi
Therefore,
\begdi
(\eta \preceq \zeta) \wedge (\zeta \preceq \gamma ) \Rightarrow \eta \preceq \gamma.
\enddi
To prove anti-symmetry, note that
\begdi
(\eta \neq \zeta) \land  (\eta \preceq \zeta) \Rightarrow \nexists \beta : \beta^{-1}(\eta) = \zeta,
\enddi
and therefore,
\begdi
(\eta \preceq \zeta) \land (\zeta \preceq \eta) \Rightarrow \eta = \zeta.
\enddi
Hence, $(\allwords,\preceq)$ is a partially ordered set.

\section{Proof of Theorem~\ref{th:dagger}}
\label{app:dagger}

A preliminary lemma is needed first.
For any fixed $\eta \in X^{\ast}$ define the {\em right concatenation map} as
\begin{equation*}
    {\cal R}_{\eta}(\zeta) = \zeta\eta, \quad \forall \zeta \in X^\ast.
\end{equation*}

\begle \label{lemma:order_pre}
Every right concatenation map is an order embedding map on $(X^\ast,\preceq)$. That is,
$\zeta \preceq \gamma$ if and only if $\zeta\eta \preceq \gamma\eta$ for all $\zeta,\gamma,\eta\in X^\ast$.
\endle

\begpr
From the definition of $\preceq$ it follows that
$$
\zeta \preceq \gamma \Longleftrightarrow \exists \lambda \in X^\ast : \gamma = \lambda\zeta.
$$
Applying ${\cal R}_\eta$ to both sides of the equality above gives
\begin{align*}
\zeta \preceq \gamma &\Longleftrightarrow \gamma \eta = \lambda\zeta\eta \\
&\Longleftrightarrow\zeta\eta\preceq \gamma\eta.
\end{align*}
\endpr

\noindent
{\bf PROOF (Theorem~\ref{th:dagger}).}
First the identity $C_i\dagger C_j=C_{i+j}$ is proved.
From the definition of the tree $C_j$ and
Lemma \ref{lemma:order_pre} it is clear that
$C_{j}\eta$ has a Hasse diagram with $\eta$ as the root and $X^j \eta$ as the set of leaf nodes.
By the definition of the dagger product, every leaf node $\eta$ of $C_i$ is replaced by $\eta C_j$
since
$$X^{i+j}=\bigsqcup\limits_{\eta \in X^i}X^j\eta.$$
Therefore, $C_i\dagger C_j$  has a Hasse diagram with $\emptyset$ as the root and $X^{i+j}$ as the set of leaf nodes, that is,
$C_i\dagger C_j=C_{i+j}$.
It is now easily checked using this identity that $(C,\dagger)$ is associative, commutative, and has $C_0$ as the unit.
Hence, $(C,\dagger)$ forms a commutative monoid. The monoid isomorphism between $\nat_0$ and $C$ is given by the bijection $i\mapsto C_i$
for all $i\in\nat_0$.
\end{document}